\documentclass[sigconf,10pt]{acmart}

\AtBeginDocument{%
  \providecommand\BibTeX{{%
    \normalfont B\kern-0.5em{\scshape i\kern-0.25em b}\kern-0.8em\TeX}}}

\usepackage[english]{babel}
\usepackage{graphicx}
\usepackage{graphics}
\usepackage[utf8]{inputenc}
\usepackage{amsmath}
\usepackage{tabularx}
\usepackage{array, multirow} 
\usepackage{subcaption}
\usepackage{xcolor}
\makeatletter

\renewcommand{\authornote}[1]{%
  \if@ACM@anonymous\else
    \g@addto@macro\@authornotes{\footnotetext{#1}}%
  \fi}
\renewcommand{\shortauthors}{Haarika Manda et al.}
\makeatother

\acmYear{2024}
\copyrightyear{2024}
\setcopyright{acmlicensed}
\acmConference[ACM SIGCOMM '24]{ACM SIGCOMM 2024 Conference}{August 4--8, 2024}{Sydney, NSW, Australia}
\acmBooktitle{ACM SIGCOMM 2024 Conference (ACM SIGCOMM '24), August 4--8, 2024, Sydney, NSW, Australia}
\acmDOI{10.1145/3651890.3672272}
\acmISBN{979-8-4007-0614-1/24/08}

\begin{document}

\title{The Efficacy of the Connect America Fund in Addressing US Internet Access Inequities}

\author{
Haarika Manda\textsuperscript{*}, 
Varshika Srinivasavaradhan\textsuperscript{*}, 
Laasya Koduru, 
Kevin Zhang, 
Xuanhe Zhou, 
Udit Paul\textsuperscript{\dag}, 
Elizabeth Belding, 
Arpit Gupta, 
Tejas N. Narechania\textsuperscript{\ddag}
}
\authornote{*Authors contributed equally to this work.}

\affiliation{%
  \institution{
UC Santa Barbara 
  \quad \textsuperscript{\dag} Ookla Inc.
    \quad \textsuperscript{\ddag} UC Berkeley}
    \country{} 
}

\ccsdesc[500]{Networks~Network measurement}
\ccsdesc[500]{Social and professional topics~Government technology policy}
\ccsdesc[500]{Social and professional topics~Broadband access}
\ccsdesc[500]{Networks~Public Internet}
\ccsdesc[500]{Social and professional topics~Governmental regulations}

\keywords{Broadband access; Connect America Fund; Internet access}

\renewcommand{\shortauthors}{}
\renewcommand{\shortauthors}{Haarika Manda et al.}
\newcommand{\smartparagraph}[1]{\noindent{\bf #1}\ }
\begin{sloppypar}
\begin{abstract}
Residential fixed broadband internet access in the US remains inequitable, despite significant taxpayer investment. This paper evaluates the efficacy of the Connect America Fund (CAF), which subsidizes new broadband monopolies in underserved areas to provide internet access comparable to that in urban regions. CAF's oversight relies heavily on self-reported data from internet service providers (ISPs). Unfortunately, the reliability of this self-reported data has always been open to question.
We use the broadband-plan querying tool (BQT) to create a novel dataset that complements ISP-reported information with ISP-advertised broadband plan details from publicly accessible websites for 537k residential addresses across 15 states. Our analysis reveals significant discrepancies, with a serviceability rate of only 55.45\%, indicating that a significant fraction of addresses certified as served are still unserved. Furthermore, we observe a compliance rate of only 33.03\%, indicating that a significant fraction of served addresses receive download speeds that are non-compliant with the FCC's 10 Mbps threshold for CAF-served addresses. Although we observe that CAF-served addresses occasionally receive higher download speeds than their monopoly-served neighbors, overall, the CAF program has largely failed to achieve its intended goal, leaving many targeted rural communities with inadequate or no broadband connectivity.

\end{abstract}

\maketitle

\vspace*{-0.2in}
\section{Introduction}
\label{sec:intro}

Broadband internet access is essential for modern civic and economic life. Medical appointments, government services, educational opportunities, and various commercial activities are all available online.  Nevertheless, 
significant inequities in broadband access persist, particularly in terms of service availability, maximum download speed, and cost.

These disparities have been the subject of persistent attention from US policymakers. The federal government has funded various programs, such as the Affordable Connectivity Program (ACP), the Connect America Fund (CAF), and the Broadband Equity, Access, and Deployment (BEAD) program, to address these inequities. Each program funds a distinct mechanism for addressing concerns regarding broadband availability, affordability, and adoption. For instance, ACP subsidizes subscription costs for qualifying consumers to make broadband more affordable. CAF, in contrast, offers subsidies to ISPs for deploying new infrastructure to improve broadband availability in underserved regions. 

Given the scale and social importance of these policy interventions, it is critical to assess their efficacy. Effective oversight of these regulatory programs would help us evaluate their success, prioritize among them, and design improvements, or new programs altogether, that better address broadband availability, affordability, and adoption concerns. An important prerequisite to any of this important policy work is to find and collect the ``right'' data for measuring and evaluating these regulatory programs.


This paper aims to take a 
step in this direction by both assessing the success of a recent regulatory program and addressing the data-related challenges involved. Specifically, we examine the extent to which the CAF program has achieved its goals. The CAF program operates by establishing a regulated monopoly in underserved communities. In locations where no provider offers service, CAF grants infrastructure subsidies to a single ISP on the condition that it agrees to serve that location. To mitigate concerns that such a subsidized monopolist may charge monopoly prices or offer substandard service, 
CAF mandates that the broadband plans offered at subsidized locations satisfy certain rate and service conditions~\cite{connect_america_fund}. Specifically, subsidized residences should have access to broadband plans similar to those offered in competitively served regions (such as urban areas) and unlike those offered by unregulated monopolists~\cite{communications_act_1996}.

To assess whether CAF achieves its  goals, we divide our analysis into three questions: (1)~Do ISPs genuinely offer broadband internet access to the addresses CAF aims to serve? (2)~Do ISPs satisfy the rate and service quality standards specified for this program? And (3)~do the regulated monopolies created through this program offer better broadband service than unregulated monopolies? That is, do the conditions placed in exchange for public subsidies through CAF curb an ISP's monopoly powers and thus offer meaningful improvements in the value that consumers receive?

To answer these three questions, we begin with an analysis of the existing public data. Specifically, CAF requires that subsidized ISPs ``certify'' the residential addresses they serve using the CAF funds, including descriptions of the broadband plans (i.e., maximum speed and price) they offer at these addresses~\cite{connect_america_fund}. In theory, one can use this self-reported data to answer the first two questions above. However, 
both federal~\cite{capito_fcc_caution} and state~\cite{ms_psc_att_subpoena} policymakers have raised concerns about the ISPs' self-reported data, and these specific complaints reflect a more general skepticism regarding the reliability of self-reported ISP data, both for CAF~\cite{gao_report} and for other broadband-related regulatory programs~\cite{wsj_charter_internet}. 
Moreover, this data alone is insufficient to answer our third policy question, because that question requires comparing CAF-related data with complementary data regarding the broadband plans advertised to nearby non-CAF addresses.\footnote{In this paper, we use the term ``CAF address'' to denote a residential address for which an ISP has been provided CAF subsidies to offer fixed broadband service. ``Non-CAF address'' refers to a residential address for which such a subsidy is not available. 
}

We leverage the broadband-plan querying tool (BQT)~\cite{Paul:Sigcomm23} to bridge this data gap. This tool mimics a real user's interaction with ISP websites to gather data on advertised broadband speeds and prices at the granularity of street addresses.
In theory, one might use BQT to query all 6+ million street addresses supported by the CAF program, as well as tens of millions of neighboring non-CAF addresses, to answer our questions. In reality, however, ethically querying addresses at that scale and in a manner that does not overwhelm the ISP's infrastructure would take more than 6 months (calculated using the median query time for each ISP). Furthermore, scaling up our data collection to increase the number of consecutive queries was found to overload the website. 
Given the dynamic nature of broadband plans and price points, such a lengthy querying time might impact our conclusions.

To address this challenge, we narrow our focus to three ISPs---AT\&T, CenturyLink, and Frontier---that receive the most significant proportion of funding from the CAF program. Additionally, we study a smaller ISP, Consolidated, for reasons highlighted in Section~\ref{ssec:sampling}. We develop a novel methodology to sample and query a subset of addresses within a census block group (CBG) without compromising the statistical significance of our results. 
We then use BQT to curate a novel dataset capturing the information (i.e., certification of service availability and offered maximum download speed) that these four ISPs report to regulators, as well as the broadband plans they publicly advertise on their websites for potential customers. Our dataset includes approximately 687~k residential addresses across 15 states, consisting of 537~k CAF addresses, identified using our sampling methodology, and 150~k non-CAF addresses. We then use this dataset to answer all three policy questions---enabling us to assess how well CAF achieves its stated aims.


Our analysis uncovers significant discrepancies between ISP-reported service data and actual broadband availability. Specifically, we observe 
that the serviceability rate---defined as the fraction of addresses ISPs actively serve out of the total queried, weighted by the number of CAF addresses in a census block group---is only 55.45\%. Worryingly, this metric drops to as low as 18\% in some AT\&T-served states. Even worse, we find that the compliance rate---defined as the weighted fraction of addresses where ISPs actively serve and advertise download speeds above the FCC's 10 Mbps threshold---is only \textbf{33.03\%}. On the positive side, we observe that when the CAF program is implemented as intended, it benefits end users in some cases. Specifically, comparing broadband plans between CAF addresses (subject to CAF's service quality and rate requirements) and unregulated non-CAF addresses in the same census block served by the same ISP, we find that CAF addresses were offered better plans 27\% of the time, with a median improvement in download speeds of 75\%. 

These results indicate that while a few users have benefited from this multi-billion dollar program, it has largely failed to achieve its intended goal, leaving many targeted rural communities with inadequate or no broadband connectivity. These findings have direct implications for policymaking. For example, as states develop plans to allocate the over \$42 billion available through the BEAD program, these results underscore the importance of independent post-hoc verification of ISP claims. More broadly, they motivate further investments in tools and infrastructure to bridge prevailing data gaps, facilitating public scrutiny of such high-stakes policy interventions.

\vspace*{-0.15in}

\section{Background \& Motivation}
\label{sec:background}
\subsection{Policy Interventions in the US}
The history of policy interventions in the US illustrates a deep-rooted commitment to ``universal service," an ideal that assures all Americans access to essential communications technologies. 
For instance, the Telecommunications Act of 1996 catalyzed the establishment of the Universal Service Fund (USF), which includes programs such as Lifeline~\cite{lifeline_program_fcc}, E-rate~\cite{erate_program_fcc}, and the Rural Health Care program~\cite{rural_health_care_program_fcc}, designed to subsidize access for various sectors of the community. The Connect America Fund (CAF)~\cite{connect_america_fund} within USF focuses on deploying infrastructure to hard-to-serve areas, with over \$10 billion allocated over seven years. Subsequent initiatives, such as the Rural Digital Opportunity Fund (RDOF)~\cite{fcc_rural_digital_opportunity_fund}, demonstrate an ongoing federal commitment to broadband availability, among other priorities.
Further advancing these efforts, the Infrastructure Investment and Jobs Act introduced the Broadband Equity, Access, and Deployment (BEAD) Program~\cite{ntia_bead_program}, directing over \$42 billion to enhance broadband availability. 

More broadly, efforts to ensure universal service extend beyond CAF, RDOF, and BEAD to include the Affordable Connectivity Program (ACP)~\cite{fcc_affordable_connectivity_program}, the Digital Equity Act~\cite{ntia_digital_equity_act_programs} and state-level universal service funds, reflecting a multifaceted approach to resolving broadband access disparities. These initiatives collectively aim to fulfill the enduring goal of universal service by improving broadband availability, affordability, and adoption across the nation.

\subsection{The Connect America Fund}
\label{sec:caf}

This paper dives deeper into the Connect America Fund (CAF) as a case study to understand how well a high-stakes policy intervention may realize its intended goal. 

\smartparagraph{Program overview.}
Before the FCC created the CAF program in 2011, the USF subsidized competitive access to traditional telephony, supporting telephone network infrastructure for multiple providers. However, with the introduction of CAF, the FCC shifted its focus to subsidizing a {\em single} provider committed to developing {\em broadband-capable} networks in high-cost regions. This shift marked a significant change in the structure of the USF's high-cost support.

CAF funding was structured into several phases, with some phases further divided. One notable phase, CAF's Phase II, was split into two distinct programs. The first, the CAF II Model, offered certain ISPs a predetermined subsidy based on the FCC's forward-looking cost model. In return for accepting the subsidy, these ISPs promised to serve locations deemed eligible for funding support by the FCC. This program had a deadline, requiring all providers receiving CAF II Model support to complete their network deployments by the end of 2021, after accounting for a one-year extension (including one additional year of funding support)~\cite{usac_caf_phase_ii}.


As noted, the FCC's rules for CAF are designed to subsidize only one provider in regions that lack broadband internet access service. That is, CAF subsidizes the creation of a monopolist~\cite{connect_america_fund}. As a result, CAF rules mandate that subsidized providers offer services that are ``reasonably comparable" to those available in competitive areas, like urban centers, and at rates ``reasonably comparable'' to those charged in such areas~\cite{communications_act_1996}. Essentially, because CAF recipients are subsidized monopolists, they are subject to specific cost and service conditions intended to curb their monopoly powers.


\begin{figure*}[t]
\centering
\begin{subfigure}[b]{0.32\linewidth}
\includegraphics[width=\textwidth]{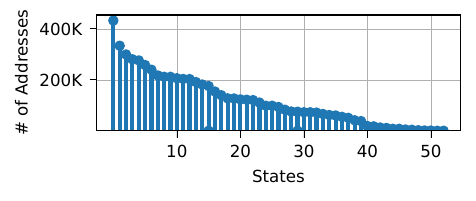}
\caption{CAF addresses across states\label{1a}} 
\end{subfigure}%
\begin{subfigure}[b]{0.32\linewidth}
\includegraphics[width=\textwidth]{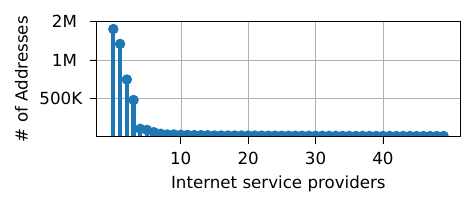}
\caption{CAF addresses across ISPs\label{1b}} 
\end{subfigure}
\begin{subfigure}[b]{0.32\linewidth}
\includegraphics[width=\textwidth]{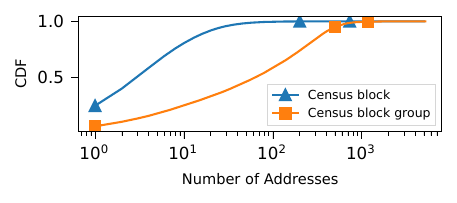}
\caption{CAF addresses across CBs and CBGs\label{1c}} 
\end{subfigure}
\begin{subfigure}[b]{0.32\linewidth}
\includegraphics[width=\textwidth]{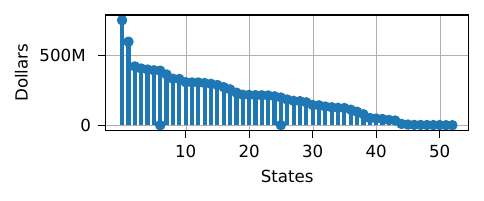}
\caption{CAF II disbursements across states\label{1d}} 
\end{subfigure}%
\begin{subfigure}[b]{0.32\linewidth}
\includegraphics[width=\textwidth]{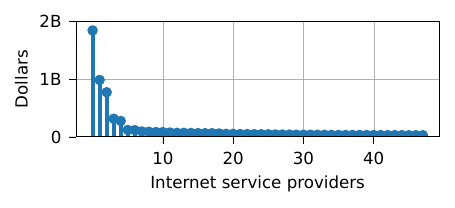}
\caption{CAF~II disbursements across ISPs\label{1e}} 
\end{subfigure}
\begin{subfigure}[b]{0.32\linewidth}
\includegraphics[width=\textwidth]{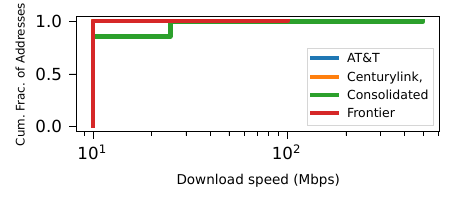}
\caption{Certified download speeds\label{1f}} 
\end{subfigure}
\caption{Attributes of the existing public CAF program datasets. 
\label{funds_disbursed}}
\end{figure*}

\smartparagraph{CAF regulations.}
ISPs that receive subsidies through CAF must satisfy certain regulatory obligations. The most basic of these is that ISPs must build and offer service within the regions they agreed to serve by accepting the CAF subsidies. The FCC considers deployment to a location complete if the carrier either offers broadband service at that location or if it could provide service within ten business days upon request~\cite{fcc2016broadband}. Moreover, that service must offer download speeds of at least 10~Mbps and upload speeds of at least 1~Mbps~\cite{fcc2014connectamerica}. Finally, the FCC specifies the maximum rates that ISPs may charge at subsidized locations. Specifically, the FCC has stated that it will deem a rate to be ``reasonably comparable'' to those available in urban regions if it is within two standard deviations of the average rate charged in urban locales for similar service, based on the FCC's annual survey of urban rates~\cite{fcc2014connectamerica}. Employing this approach, the FCC set a price cap for 2024 of approximately \$89 per month for 10/1~Mbps service~\cite{fccpublicnotice}. 

\smartparagraph{Regulatory oversight.}
Policymakers have recognized the importance of evaluating the effectiveness of such a billion-dollar program in achieving its intended goal, and so have implemented some mechanisms for regulatory oversight. These oversight mechanisms focus on two key questions: (1)~{\em service availability}, namely, whether ISPs are using CAF funds to provide internet service to those consumers the program is intended to help, and (2)~{\em service compliance}, namely, whether those consumers are receiving service that complies with the FCC's rate and service (i.e., download speed) standards. To address these questions, the FCC has entrusted the {\em Universal Service Administrative Company (USAC)}, an independent not-for-profit entity, with both the management of CAF funds and the duty of monitoring compliance with specific CAF conditions, including deployment milestones, and rate and service standards. 

To answer the first question, USAC requires ISPs to certify deployment progress annually. It verifies these certifications by requiring documentary evidence from ISPs proving their ability to offer service at the certified locations. Acceptable documentation for this certification process includes screenshots of a public-facing availability tool (i.e., the ISP's website), subscriber bills, or internal emails from engineering divisions authorizing the release of those locations to sales and marketing teams. For the second question, USAC requires performance testing to ensure that the broadband internet access service subsidized by CAF meets speed and latency standards. ISPs are required to conduct these tests from the premises of active subscribers to a remote test server at an FCC-designated Internet exchange point.

\vspace*{-0.2in}
\subsection{Dataset Available to Policymakers}


In its open data repository, USAC offers a public dataset~\cite{usac_dataset} of locations subsidized by USF programs, specifically focusing on the CAF Map dataset~\cite{usac_caf_map_dataset}. Derived from carriers' annual reports via the {High-Cost Universal Broadband (HUBB) portal~\cite{HUBB}}, this dataset records the ISPs' certified deployment data. As noted, USAC annually verifies a random sample of these locations to check compliance with CAF obligations.
Our analysis targets addresses supported by the CAF~II~Model program within this dataset, chosen for being the largest and only fully deployed subset of USAC's High-Cost Dataset. With the deployment deadline surpassed (end of 2022, including extensions), the expectation is that ISPs are compliant with the FCC's requirements. That is, broadband is not only available (within ten days) at the ISP-certified locations in this USAC dataset but also the offered broadband plans comply with the FCC's service (at least 10~Mbps download and 1~Mbps upload speeds) and rate (at a maximum of \$89/month) requirements for the CAF program.

This USAC dataset encompasses detailed deployment information, listing each residential address with identifiers including street address, geographical coordinates, census block (CB), ZIP code, and state. It also includes the number of households certified as served, certifying ISP's name, last-mile connectivity technology, and service quality metrics (upload/download speeds and latency), along with state-wise disbursed funds.  The dataset includes 6.13~M deployment locations across the US, disbursing around \textit{\$~10~billion} to 819 different ISPs. These locations cover 787~k census blocks (6~\% of the 11~M total nationwide) and 43~k census block groups (20~\% of the 217~k total). The CAF program predominantly targets rural areas; 96.7~\% of CAF census blocks are rural, covering 26.5~\% of all US rural census blocks (2.73~M). 

We illustrate some of the key attributes of the CAF program in Figure~\ref{funds_disbursed}.  Specifically, 
Figures~\ref{1a} and \ref{1d} show the state-wise distribution of CAF addresses and disbursed funds, respectively, highlighting a concentration in the top-20 states, which account for over 73~\% of addresses. The top three states by number of CAF addresses are {Texas, Wisconsin, and Minnesota}, and by disbursement amounts are {Texas, Minnesota, and Arkansas}. Figure~\ref{1b} details the distribution of addresses certified as served by CAF-funded ISPs, while Figure~\ref{1e} shows fund distribution across ISPs. The top-4 ISPs (AT\&T, CenturyLink, Frontier, Windstream) serve 62~\% of addresses and receive 37.5~\% of the funds. Figure~\ref{1c} reveals a high variance in CAF addresses per census block, with a range from 1 to over 5~k addresses. For census block groups, the min, median, and max numbers of addresses are {1, 64, and 5.2~k}, respectively. With few exceptions, almost all ISPs certify offering a download speed of 10~Mbps (Figure~\ref{1f}), which satisfies FCC's service quality requirements.

\vspace*{-0.2in}
\subsection{The Limits of Existing Oversight}
While the USAC regulatory oversight is designed to ensure the effectiveness of the CAF program, there are significant limitations. Despite making ISPs' self-reported certifications public and validating these against CAF's service area criteria, critical gaps in compliance and oversight remain. For instance, while USAC verifies that ISPs serve certified locations and reports a ``compliance gap" metric, details about this gap are scarce. This includes uncertainty about the geographic distribution of non-compliance, whether it disproportionately affects certain populations, and whether non-compliance is due to a total lack of service or inadequate service that fails to meet CAF's standards~\cite{usac2024fccplan}.

Moreover, USAC's verification process lacks transparency, particularly regarding service availability, and some of its tests are confined to active subscribers. This limitation hinders the ability to determine if all certified locations truly receive compliant service~\cite{usac_high_cost_reviews}. The reliability of ISPs' self-reported data to USAC and the adequacy of USAC's oversight have been questioned by various stakeholders, including federal~\cite{capito_fcc_caution, gao_report} and state officials~\cite{ms_psc_att_subpoena}. For example, an investigative subpoena was issued to AT\&T over concerns about its reported service to 133~k locations, highlighting the mistrust in ISPs' self-reporting and the need for more rigorous verification~\cite{ms_psc_att_subpoena}. These insufficiencies in the existing oversight framework highlight the need for a more exhaustive and transparent evaluation framework to assess the CAF's efficacy in realizing its intended goals. 

\vspace*{-0.1in}
\section{Augmenting the CAF Dataset}
\label{sec:augment}
In this section, we describe our approach to augmenting the existing CAF dataset to address the three policy questions outlined in Section~\ref{sec:intro}. 

\vspace*{-0.1in}
\subsection{Address Selection Strategy}
\label{ssec:sampling}
To effectively answer the first two questions, our approach involves auditing and enhancing the ISPs' self-reported data with the specific plan information (if any) they advertise to consumers on their websites. For the third question, which aims to assess the efficacy of the CAF's strategy of regulating subsidized monopolists, we expand our view of the CAF dataset with additional data points. More precisely, we include new street addresses within the same census blocks as CAF-served addresses but not covered by the CAF program; we call these addresses ``non-CAF addresses''. This added information is crucial to compare each ISP's service at CAF addresses with its service in neighboring non-CAF addresses, where the ISP either functions as an unregulated monopolist or faces competition from other providers.

\smartparagraph{Why do we need to sample addresses?}
In an ideal scenario, the expansion of the dataset would involve adding all non-CAF addresses within CAF-served census blocks, followed by querying the broadband plans for every address in this augmented dataset. However, this data-collection strategy faces practical challenges in terms of scale. Conducting queries for broadband plans at such a magnitude, without overburdening the ISPs' infrastructure, would be a year-long endeavor based on our current query times. 
Consequently, our analysis is focused on a subset of street addresses from the CAF dataset within selected states and ISPs. Within each state, for every ISP in our study, we carefully select a representative subset of both CAF and non-CAF addresses in a census block. The subsequent sections detail our criteria for choosing this subset of states and ISPs and our methodology for selecting the addresses within each state for each ISP.

\smartparagraph{How do we select a subset of ISPs?}
Given the skewed distribution of fund disbursements across ISPs, where only the top few received the majority of the funds, we consider the top three ISPs: AT\&T, CenturyLink, and Frontier. Collectively, these ISPs received 37.5\% of the total \$10 billion in funding to serve 54\% of the total 6.13~M CAF addresses, spanning 43 US states. Additionally, to contrast with these top three ISPs, we also study Consolidated Communications, a comparatively smaller player in the CAF program.
Consolidated Communications received \$193~M CAF funding to provide broadband connectivity to 138~k addresses, covering 16.2~k census blocks, 1.3~k census block groups, and 21 states. Despite certifying service to only 18\% of the number of addresses certified by our third top ISP (Frontier), Consolidated Communications ranks fifth in terms of the number of addresses served through the CAF program.

\smartparagraph{How do we select a subset of states?}
The distribution of addresses certified as served or funds disbursed by the state is not as skewed as that of the ISPs. Thus selecting a subset of states for analysis is less straightforward. We chose a subset of 15 states that capture a range of attributes. Specifically, we considered a mix of states where our selected providers were dominant ISPs, serving more than 80\% of the CAF addresses. For instance, AT\&T serves 94\% of addresses in Mississippi and 82\% in Georgia. We also selected states where multiple of our chosen providers certified service to nearly equal numbers of addresses, such as Wisconsin (CenturyLink and Frontier) and California (AT\&T and Frontier). 
We also ensured that we had states spanning major US geographic regions, as well as representing a wide distribution of sizes, from the most populous (California) to one of the least populous (Vermont). This mix provides a broad sample of the national ISP landscape and enables a nuanced analysis 
across diverse regional contexts.

\smartparagraph{How do we select street addresses?}
Next, our objective is to determine which street addresses to query for a given ISP in a specific state. One possible approach is to randomly sample CAF addresses across the state. However, due to the uneven distribution of CAF addresses across different census blocks and block groups, this method might over-sample from a limited number of census blocks or block groups. Such sampling might not accurately reflect the implementation of the CAF program across various regions within the state, each with distinct socioeconomic and demographic attributes. Considering that a CBG typically represents 600-3000 people with relatively homogeneous demographic and socioeconomic characteristics, our strategy aims to draw a representative number of samples from each block group. This ensures that the sampled dataset adequately captures the state's socioeconomic composition. Specifically, we aim to randomly sample at least 10\% of street addresses from the CAF dataset for each CBG. For CBGs with fewer than 300 CAF addresses (83\% as indicated in Figure~\ref{1c}), we strive to sample a minimum of 30 addresses. This approach aims to guarantee a minimum of thirty samples from each CBG for each ISP, which is crucial for the statistical significance of computed metrics such as median download speed and average carriage value. In cases where the CBG contained more than 30 addresses, we aim to sample the greater of 30 and 10\% of the addresses in the block group. In cases where CBGs have fewer than 30 CAF addresses (38\% as shown in Figure~\ref{1c}), we strive to query all addresses. 

Appendix Table~\ref{tab:data} summarizes the attributes of the data points (i.e., CAF addresses) across different states, ISPs, and CBGs that we queried after implementing all these steps. Table~\ref{tab:q3} breaks down the CAF and non-CAF addresses collected specifically for answering question 3. The methodology for this is described in Section~\ref{sec:q3}.

\vspace*{-0.15in}
\subsection{Data Collection using BQT}
\label{ssec:bqt}
Our approach utilizes and enhances the \textit{broadband-plan querying tool (BQT)~\cite{Paul:Sigcomm23}} to gather broadband plan information (speed and price data) for selected ISPs at any  US street address. As described in~\cite{Paul:Sigcomm23}, BQT inputs a street-level address and returns the set of broadband plans (upload/download speeds and corresponding prices in US dollars) offered by an ISP at that location. Leveraging BQT, we utilize The Bright Initiative's proxy service\footnote{https://brightinitiative.com/}, specifically their pool of data center and residential IP addresses,  to ensure that ISP websites receive our queries originating from a geographically diverse pool of IP addresses. By simulating the behavior of a real user interacting with the ISP website from a diverse set of endpoints, BQT facilitates querying advertised broadband plans at scale for many Docker Containers. 

Our study adopts the workflow detailed in~\cite{Paul:Sigcomm23} to query advertised broadband plans for street addresses identified in the sampling phase (detailed in Section~\ref{ssec:datacollectionworkflow}). 
The success rate of BQT in querying street addresses varies among ISPs~\cite{Paul:Sigcomm23}, with lower hit rates attributable to factors like inaccurate address inputs, updates to ISP websites including user interface (UI) changes, and enhanced bot-detection mechanisms. 
However, thanks to BQT's modular design, addressing static issues like UI changes is straightforward. To circumvent dynamic issues such as bot-detection updates, we rerun failed queries multiple times and rotate through the pool of IP addresses provided by our proxy service. For instance, if a query fails multiple times for a specific address, we select a new address from the same census block group. 

Accurate identification of unserved addresses is crucial for our analysis. Appendix Section~\ref{ssec:datacollectionworkflow} delves into the workflow we employed to compile our dataset, specifically how we addressed various ISP-specific ambiguities to distinguish between queries that failed for unknown reasons and those that failed because the addresses are unserved by an ISP.

\subsection{Ethical Considerations}
\label{ssec_ethics}
Our methodology queries ISP plans at the street address level without collecting or analyzing Personally Identifiable Information (PII). Furthermore, we utilize a private dataset obtained from Zillow under a data use agreement for the non-CAF addresses. This dataset does not disclose any individual identities, consequently, our research does not fall under the category of human subjects research. Similarly, data extracted from ISP websites are devoid of PII, ensuring we cannot identify residents, their broadband subscriptions, or the actual service performance at any given address. Our approach, aligning with previous work~\cite{Paul:Sigcomm23}, focuses on gathering information about ISP plans from their websites, a practice that adheres to both legal standards and the ethical norms of the research community~\cite{van_buren_states,hiq_linkedin,dept_justice}.


\begin{figure*}[t]
\begin{subfigure}[b]{0.33\linewidth}
   \includegraphics[width=1\linewidth]{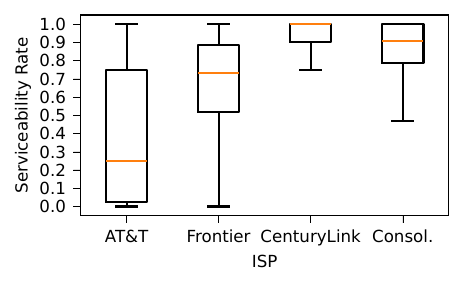}
    \caption{ISPs aggregated over the 15 states 
    }
    \label{fig:isps}
\end{subfigure}
\begin{subfigure}[b]{0.33\linewidth}
    \includegraphics[width=1\linewidth]{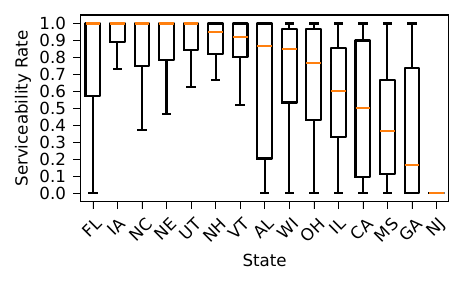}
    \caption{States aggregated over the 4 ISPs}
    \label{fig:states}
\end{subfigure}
\begin{subfigure}[b]{0.33\linewidth}
    \includegraphics[width=\linewidth]{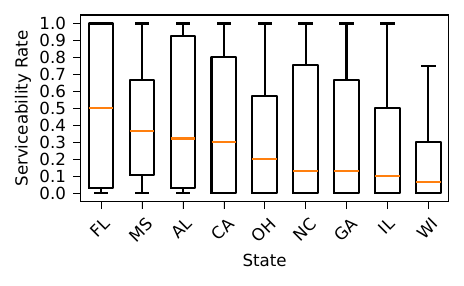}
    \caption{AT\&T across the states it serves}
    \label{fig:att}
\end{subfigure}
\caption{Serviceability rates by ISP and state.}
\label{fig:q1_state_and_isp}
\end{figure*}
\section{CAF Efficacy Analysis}
In this section, we demonstrate how our dataset enables us to answer the three key policy questions to assess the efficacy of the CAF program in achieving its intended goal.  Specifically, our goal is to assess whether the addresses certified by the ISPs are truly experiencing improvement in service availability (Section~\ref{sec:q1}), whether the services offered through the CAF program meet the minimum service quality requirements (Section~\ref{sec:q2}), and whether the creation of regulated monopolies offers better service quality to users compared to existing unregulated monopolies (Section~\ref{sec:q3}).

 \vspace*{-0.1in}

\subsection{Assessing Service Availability (Q1)}
\label{sec:q1}
We begin with an analysis of whether the CAF program has been successful in making internet service available in the underserved communities it targets. More specifically, {\em we explore whether the service is genuinely available to addresses certified as served by the ISPs through the CAF program}. To address this question, we calculate the \textit{serviceability rate}, a metric we define to quantify the fraction of actively served addresses. When reporting serviceability rates at the spatial granularity of a census block group (CBG), we compute the fraction of addresses served over the addresses successfully queried. When reporting results at coarser granularities (e.g., for a given state or group of states), we weight the serviceability rate at the block group level with the total number of CAF addresses for the CBG. We employ this weighted approach to ensure that our sampling strategy, which varies for census block groups of different population sizes, does not skew our aggregated results. Ideally, this metric should be close to 100\%, indicating that ISPs serve all the addresses they certify as served in the USAC dataset.



\smartparagraph{The big picture.}
We queried 409,268 street addresses across 15 states, each of which was claimed to be served by at least one of the four ISPs in our study, to answer this question. Aggregated across the states, we observed a serviceability rate of 55.45\%, meaning that we estimate only 55.45\% of the addresses in these states, certified as served by these ISPs, are indeed served by these ISPs. This finding is concerning as it suggests that the internet service is still ``unavailable'' for a significant fraction of addresses certified as served in the USAC dataset.   We note that we could not ascertain the presence or absence of service availability in 0.96\% of addresses, where our query failed for reasons related to address resolution. We elaborate on these limitations in Section~\ref{sec:limitations}. 

\smartparagraph{Service availability across ISPs.}
To examine how the serviceability rate varies across different ISPs, we disaggregated the data by ISPs. We observed serviceability rates of 31.53\% , 70.71\% , 90.42\% , and 83.95\% for AT\&T, 
Frontier, CenturyLink, and Consolidated, respectively. 
The fact that CenturyLink has the highest serviceability rate is encouraging because they received the most funding (\${1.84} billion) of any ISP through the CAF program. At the same time, it is concerning to see such low serviceability rates for AT\&T and Frontier, considering that they rank second and third among all ISPs in terms of total funds received.

To gain further insight into the spatial distribution of this metric, we evaluated the serviceability rate at the granularity of census block groups (Figure~\ref{fig:isps}). This analysis reveals that although AT\&T's serviceability rate is the lowest, the other ISPs also exhibit low serviceability rates in a significant fraction of census block groups. More specifically, we observe that the serviceability rate for the lower quartile (i.e., bottom 25\%) is only 53\% for Frontier and 78\% for Consolidated.  

\smartparagraph{Service availability across states.}
We next examine whether the low serviceability is skewed towards a few states or evenly spread across all. Figure~\ref{fig:states} shows the distribution of serviceability rates across different census block groups for all fifteen states. We observe that some regions in a state are more affected than others. For example, the median serviceability across all CBGs is very high in Florida (FL) and Alabama (AL), but it is quite low for the lower quartiles. We also observe that serviceability rates are low across a significant fraction of block groups in New Jersey (NJ), Georgia (GA), Missouri (MS), and California (CA). 


\smartparagraph{Service availability across state-ISP pairs.}
We next explore whether the low serviceability rates for specific states are attributable to a few specific large ISPs. To do so, we disaggregate the serviceability rate data by state and ISP pair. We observe low variability in serviceability rates across states, except for CenturyLink in New Jersey and Frontier in Florida.  In these two states,  serviceability rates significantly diverge from those in other states served by these ISPs. Especially surprising was the 0\% serviceability rate of  Centurylink in New Jersey for 980 queried  addresses. Further, we note that states with low serviceability rates are predominantly served by AT\&T. Figure~\ref{fig:att} demonstrates that AT\&T consistently exhibits low serviceability rates across all nine states.

\begin{figure}[t!]
\begin{subfigure}[b]{.48\linewidth}
    \includegraphics[width=\linewidth]{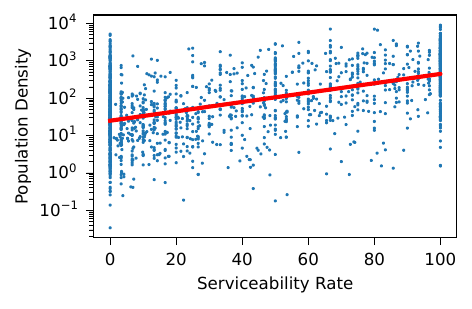}
    \caption{California}
    \label{fig:popden_ca}
\end{subfigure}
\begin{subfigure}[b]{.48\linewidth}
    \includegraphics[width=\linewidth]{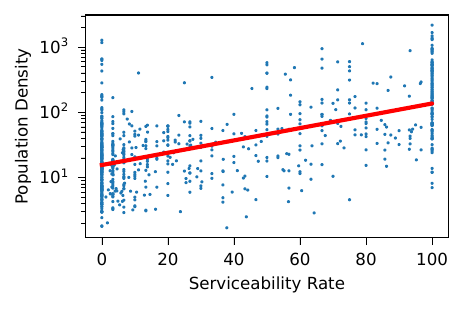}
    \caption{Georgia}
    \label{fig:popden_ga}
\end{subfigure}
\caption{Population density (people/mi$^2$) vs. AT\&T serviceability rates for two high-population states.}
\label{fig:q1_deepdive}
\vspace*{-0.2in}
\end{figure}

\begin{table*}[t]
  \centering
  \scriptsize
  \caption{Differences in the certified (from USAC) and advertised (from BQT) maximum download speeds (Mbps). All non-compliant plans are highlighted in \textbf{\textcolor{red}{red}}. The numbers in parentheses indicate the total CAF addresses of the ISP. ``0 Mbps" entries indicate the percent of addresses that are unserved.} 
  \label{tab:q2}
    \resizebox{\textwidth}{!}{%
  \begin{tabular}{cccc|cccc|cccc|cccc}
    \multicolumn{4}{c}{\textbf{AT\&T (176.53k)}} & 
    \multicolumn{4}{c}{\textbf{CenturyLink (82.83k)}} & 
    \multicolumn{4}{c}{\textbf{Consolidated (21.98k)}} & 
    \multicolumn{4}{c}{\textbf{Frontier (127.86k)}} \\ \hline
    \multicolumn{2}{c}{\textbf{Certified}} & 
    \multicolumn{2}{c|}{\textbf{Advertised}} & 
    \multicolumn{2}{c}{\textbf{Certified}} & 
    \multicolumn{2}{c|}{\textbf{Advertised}} & 
    \multicolumn{2}{c}{\textbf{Certified}} & 
    \multicolumn{2}{c|}{\textbf{Advertised}} & 
    \multicolumn{2}{c}{\textbf{Certified}} & 
    \multicolumn{2}{c}{\textbf{Advertised}} \\
     {\textbf{Mbps}} & \textbf{\%} & 
     {\textbf{Mbps}} & \textbf{\%} & 
     {\textbf{Mbps}} & \textbf{\%} & 
     {\textbf{Mbps}} & \textbf{\%} &
     {\textbf{Mbps}} & \textbf{\%} & 
     {\textbf{Mbps}} & \textbf{\%} & 
     {\textbf{Mbps}} & \textbf{\%} & 
     {\textbf{Mbps}} & \textbf{\%}
     \\ \hline\hline
   
10 & 100 & \textbf{\textcolor{red}{0}} & \textbf{\textcolor{red}{67.660}} & 10 & 100 & \textbf{\textcolor{red}{0}} & \textbf{\textcolor{red}{10.024}} & 10 & 86.02 & \textbf{\textcolor{red}{0}} & \textbf{\textcolor{red}{14.543}} & 10 & 99.98 & \textbf{\textcolor{red}{0}} & \textbf{\textcolor{red}{30.611}} \\
 &  & \textbf{\textcolor{red}{AT\&T Internet Air}} & \textbf{\textcolor{red}{5.052}} &  &  & \textbf{\textcolor{red}{0.5}} & \textbf{\textcolor{red}{0.298}} & 25 & 12.87 & \textbf{\textcolor{red}{3}} & \textbf{\textcolor{red}{0.027}} & 100 & 0.01 & \textbf{\textcolor{red}{Frontier Internet}} & \textbf{\textcolor{red}{53.255}} \\
 &  & \textbf{\textcolor{red}{0.768}} & \textbf{\textcolor{red}{1.153}} &  &  & \textbf{\textcolor{red}{1.5}} & \textbf{\textcolor{red}{1.996}} & 100 & 0.77 & \textbf{\textcolor{red}{7}} & \textbf{\textcolor{red}{0.177}} &  &  & Unknown Plan & 12.138 \\
 &  & \textbf{\textcolor{red}{1}} & \textbf{\textcolor{red}{0.976}} &  &  & \textbf{\textcolor{red}{3}} & \textbf{\textcolor{red}{15.036}} & 1000 & 0.33 & 10 & 12.477 &  &  & 10 & 0 \\
 &  & \textbf{\textcolor{red}{3}} & \textbf{\textcolor{red}{1.786}} &  &  & \textbf{\textcolor{red}{6}} & \textbf{\textcolor{red}{5.664}} &  &  & 11-99 & 42.323 &  &  & 11-99 & 0 \\
 &  & \textbf{\textcolor{red}{5}} & \textbf{\textcolor{red}{2.479}} &  &  & 10 & 32.52 &  &  & 100-999 & 1.159 &  &  & 100-999 & 0.098 \\
 &  & 10 & 3.135 &  &  & 11-99 & 34.145 &  &  & 1000+ & 29.295 &  &  & 1000+ & 3.895 \\
 &  & 11-99 & 9.628 &  &  & 100-999 & 1.78 &  &  &  &  &  &  &  &  \\
 &  & 100-999 & 0.359 &  &  & 1000+ & 0 &  &  &  &  &  &  &  &  \\
 &  & 1000+ & 7.767 &  &  &  &  &  &  &  &  &  &  &  &  \\
    
  \end{tabular}}
\end{table*}

\smartparagraph{Who is (not) receiving service through this program?}
As was shown in Figure~\ref{fig:att}, there is significant variability in AT\&T's serviceability rates across the nine states it serves, prompting an investigation into which regions within a state experience lower rates. Our visualization of the geospatial distribution of serviceability rates reveals that, in most states, areas further from large city centers tend to have lower serviceability rates. For instance, in California (see Figure~\ref{fig:heatmap_ca} in the Appendix), areas distant from major coastal city centers exhibit lower rates. A similar pattern emerges in Georgia (see Figure~\ref{fig:heatmap_ga} in the Appendix). 

These findings led us to examine the correlation between serviceability rates and population density within a region, uncovering a strong correlation across all states served by AT\&T. For the sake of brevity, we showcase these results for two states. Figure~\ref{fig:q1_deepdive} presents these findings for California and Georgia. These results suggest that though CAF's primary targets are in underserved high-cost rural areas, the program's funds are instead often directed to regions near areas that already receive service. One hypothesis is that these areas are most probably easier to bring service to, with fewer geographic or economic barriers. We observed a similar trend in every other state---Alabama, Florida, Illinois, North Carolina, Ohio, and Wisconsin---except Mississippi, where we observed no significant correlation between serviceability rates and population density. 

\smartparagraph{Sensitivity analysis.}
Given the importance of our findings, we examine the sensitivity of our results to our sampling strategy. Our results indicate that the reported serviceability rates are robust to changes in sampling rates, and we provide those details in Section~\ref{ssec:sensitivity} of the Appendix. This analysis also underscores the diminishing returns of querying additional CAF addresses within a census block group.

\smartparagraph{Takeaways.}
Our results indicate that among the ISPs that receive the most funds through the CAF program, a significant fraction of addresses they certify as served to regulators remain unserved. Among the four ISPs we examined, AT\&T exhibits the lowest serviceability rates. Furthermore, the locations AT\&T serves under CAF are predominantly in densely populated regions, closer to urban centers. This suggests that CAF has not fully achieved its goal of servicing such high-cost areas as sparsely populated rural regions.

\vspace*{-0.1in}
\subsection{Assessing Compliance (Q2)}
\label{sec:q2}
Our next goal is to assess whether consumers are offered broadband plans from  ISPs that meet the FCC's minimum rate (price) and service quality standards set for the CAF program. To answer this question, we calculate the \textit{compliance rate}---a metric we define to represent the fraction of actively served addresses that comply with the FCC's rate (a maximum of \$89 per month) and quality requirements (a minimum of 10~Mbps download and 1~Mbps upload speeds). We use the same methodology as for the serviceability rate to report this metric at coarser granularities.



\smartparagraph{Assessing compliance with rate requirements.}
Our analysis shows that the prices charged by ISPs are below the FCC's price caps, indicating that ISPs meet the FCC's pricing requirements. Specifically, the prices offered by our analyzed ISPs, for the tier of 10~Mbps (download), ranged from \$30 to \$55 per month, which is less than the benchmark of \$89/month, as specified by the FCC. We observed similar findings for the other broadband speed tiers. This is a positive finding, suggesting compliance with regulatory standards. However, it is important to note that the FCC's rate requirements are relatively lenient. The FCC considers a rate ``reasonably comparable'' to urban prices if it falls within two standard deviations of the average urban rate. This criterion results in requiring carriage values\footnote{Carriage value is defined as the Mbps of user traffic that an ISP advertises to carry for one dollar per month~\cite{Paul:Sigcomm23, narechania2022convergence}.} of only 0.1 for 10~Mbps plans, compared to the median carriage values of 15 in competitive urban centers and 10 in non-competitive urban areas reported in previous research~\cite{Paul:Sigcomm23}---a significant difference. Nonetheless, ISPs do adhere to the FCC's pricing requirements and, in some instances, offer rates more aligned with those observed in urban markets by prior work.

\smartparagraph{Assessing compliance with service quality requirements.}
In contrast to price, our analysis of service quality, quantified in terms of download speeds\footnote{We restricted our analysis to download speeds as it exhibits higher entropy than upload speed, and because many ISPs only advertise download speeds on their websites.}, tells a very different story. We observe an aggregate compliance rate of 33.03\%. Table~\ref{tab:q2} shows the distribution of the maximum speed tiers the four ISPs advertise on their websites to all the CAF households we queried. Here, we mark the advertised speed as 0 for the unserved addresses to calculate the proportion that each speed tier accounts for in the total pool of addresses. The table also shows the certified speed tiers, i.e., the broadband speeds reported by these ISPs to USAC.


We observe the discrepancy between the broadband plans ISPs certify to regulatory bodies and what they advertise to consumers on their websites. All  speed tiers reported by ISPs to USAC exceed 10~Mbps, the FCC's minimum speed requirement for CAF-funded addresses. For instance, AT\&T has certified a 10~Mbps download speed for each of its 176.53~k CAF street addresses. However, it advertises a wide range of plans to consumers living at CAF addresses, ranging from as low as 768~Kbps to as high as 5~Gbps. We make similar observations for CenturyLink, while Consolidated and Frontier show more variation in both self-reported and consumer-advertised plans. We observe compliance rates of 16.58\%, 69.30\%, 15\%, and 85.56\% for AT\&T, CenturyLink, Frontier, and Consolidated, respectively. Although a compliance rate less than 100\% for any ISP is problematic, the low compliance rates for AT\&T and Frontier are especially concerning.


Note that in the table approximately 12\% of Frontier addresses that we queried are categorized as ``Unknown Plan'' because these addresses were active subscribers of Frontier, but Frontier's website does not display available speed tiers for such addresses. Also, we classify Frontier's ``Frontier Internet'' and AT\&T's ``Internet Air'' plan as non-compliant because neither ISP offers minimum speed guarantees for these plans. For example, Frontier explicitly states, as shown in {Appendix Figure~\ref{fig:frontier_plan_information}}, that ``\textit{Frontier Internet Service is not provided based on speed tiers or other level of performance, and Frontier does not guarantee that you will be able to perform any particular Internet activity with the service.''} The ambiguity in advertised service quality on Frontier's website could be attributed to previously settled allegations that it fraudulently exaggerated the quality of its broadband offerings~\cite{frontiercommunications2022}.

\smartparagraph{Takeaways.}
Our analysis shows an aggregate compliance rate of {33.03\%}, which is quite low. This is even more pressing given that almost all the CAF-funded addresses are situated in rural areas, which typically have limited options for broadband access. The compliance rates of around 17\% and 15\% for AT\&T and Frontier, respectively, are alarming. These findings indicate that despite benefiting financially from the CAF program, these ISPs do not comply with the FCC's service availability and quality terms, leaving many targeted rural communities with no or substandard Internet connectivity.

\begin{figure*}[t]
\centering

\begin{subfigure}[b]{0.28\linewidth}
\includegraphics[width=\textwidth]{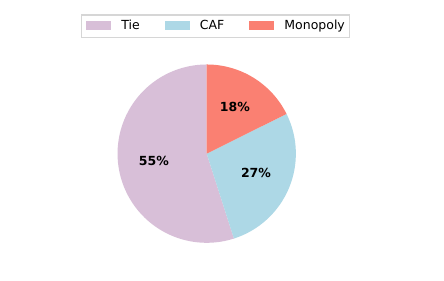}
 \caption{Composition of Type~A blocks, highlighting best-performing mode} 
\label{fig:q3_pie_1} 
\end{subfigure}%
\begin{subfigure}[b]{0.28\linewidth}
\includegraphics[width=\textwidth]{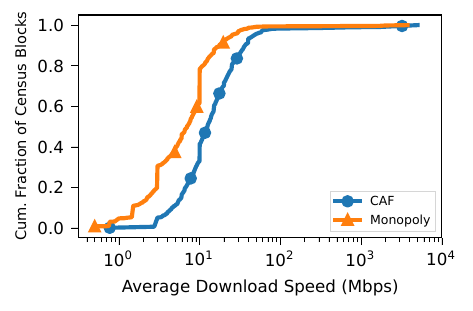}
\caption{CAF vs monopoly speeds}
\label{fig:caf_mono_speeds}
\end{subfigure}
\begin{subfigure}[b]{0.28\linewidth}
\includegraphics[width=\textwidth]{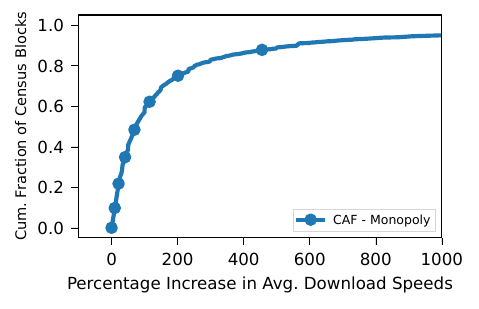}
\caption{Difference in maximum speeds}
\label{fig:caf_mono_percentage}
\end{subfigure}
\caption{Comparison of regulated monopolies
(CAF) in Type A (CAF+Monopoly) census blocks.
\label{fig:q3_caf_mono}}
\end{figure*}
\begin{figure*}[t]
\centering

\begin{subfigure}[b]{0.28\linewidth}
\includegraphics[width=\textwidth]{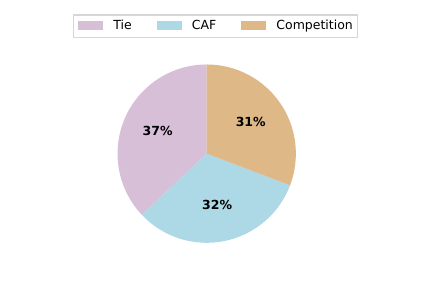}
 \caption{Composition of Type~B blocks, highlighting best-performing mode} 
\label{fig:q3_pie_2} 
\end{subfigure}%
\begin{subfigure}[b]{0.28\linewidth}
\includegraphics[width=\textwidth]{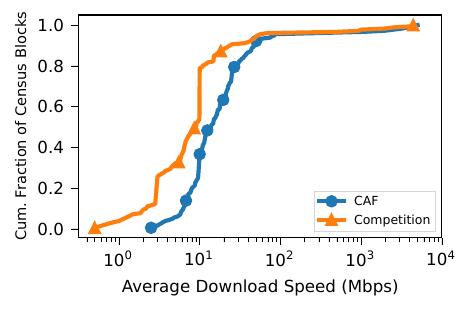}
\caption{CAF vs competition speeds}
\label{fig:caf_comp_speeds}
\end{subfigure}
\begin{subfigure}[b]{0.28\linewidth}
\includegraphics[width=\textwidth]{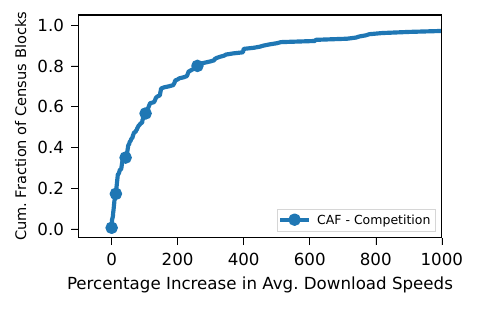}
\caption{Difference in maximum speeds}
\label{fig:caf_comp_percentage}
\end{subfigure}
\caption{Comparison of regulated monopolies
(CAF) in Type B (CAF+Competition) census blocks.
\label{fig:q3_caf_comp}}
\end{figure*}


\vspace*{-0.1in}
\subsection{Assessing Regulated Monopolies (Q3)}
\label{sec:q3}
The CAF program creates regulated monopolies by funding ISPs to be the sole broadband providers in certain areas, setting standards for service rates and quality similar to those in competitive urban markets. In this section, we examine whether CAF users served by these monopolies receive better service than those served by unregulated monopolies; and whether the available service is on par with that in locations with multiple providers.


\smartparagraph{Additional data collection.}
To answer this third policy question, we need to quantify the efficacy of CAF regulations in ensuring that subsidized locations receive Internet service at rates and speeds that are better than those offered by an unregulated monopolist, and that approximate those available in competitive markets. Such an analysis requires us to compare the available plans for CAF addresses with those of their neighbors where the incumbent ISP (i.e., the one that received the CAF funding) is operating as an unregulated monopoly, as well as where the incumbent ISP is competing with other providers. Given that census block groups could be widely spread in rural areas, we focus on addresses in the same census block, which is a much smaller geographic unit than a census block group. Because this approach requires querying considerably more addresses, we focus on a reduced subset of states (California, Utah, Illinois, Ohio, North Carolina, New Hampshire, and Georgia) while still ensuring geographic diversity. 

We apply a filter to identify census blocks served exclusively by the six ISPs—AT\&T, Frontier, CenturyLink, Comcast Xfinity, Consolidated, and Spectrum (formerly known as Charter)—currently supported by BQT. We use the data available from FCC Form 477 and the National Broadband Map datasets for this step; the result is 20.8~k census blocks. For these census blocks, we enumerate all CAF addresses from the USAC dataset and non-CAF addresses from a dataset of residential addresses provided by Zillow~\cite{zillow2023}. This results in a list of 216~k CAF and 150~k non-CAF addresses to query. We query each of these addresses using BQT and then filter out the ones that are not serviceable(see Table~\ref{tab:q3} for details). We also filter out census blocks where we do not find any non-CAF address served by the CAF-funded ISP. After each of these filtering steps, the result is 90~k CAF and 60~k non-CAF addresses across 9.4~k census blocks for our analysis of question three. We emphasize that this study is multiple orders of magnitude greater than any previous attempt to quantify the efficacy of CAF's rate and service conditions~\cite{narechania2022convergence}.

\smartparagraph{Methodology.}
We first consider three modes in which an incumbent ISP, funded through the CAF program, operates within a census block: \textit{regulated monopoly} for addresses supported through the CAF program, \textit{unregulated monopoly} for non-CAF addresses where it is the sole service provider, and \textit{competition} for non-CAF addresses where it faces competition from one or more service providers. For clarity, we will refer to regulated monopolies as CAF and unregulated monopolies simply as monopolies. As previously mentioned, we treat addresses within the same census block as neighbors since they share various geospatial characteristics. Given these modes and choices of spatial granularity, we consider three different types of CAF-served census blocks: \textit{Type~A}, where the CAF ISP is only operating in CAF and monopoly modes; \textit{Type~B}, where the CAF ISP is only operating in CAF and competition modes; and \textit{Type~C}, where the CAF ISP is operating in all three modes. In our dataset, we have 8.76~k, 0.56~k, and 0.10~k census blocks for each of these categories, respectively. It is not surprising that we do not see a lot of competition for CAF-served census blocks, as they are predominantly in rural communities that are typically served, if at all, by a single provider.

To compare plans advertised by the CAF ISP in different modes, we compute the average of a metric that indicates service quality, such as maximum download speed, price, maximum carriage value, etc., and report the difference in average values at census block granularity. For simplicity, we use \textit{download speed} to denote the maximum download speed an ISP offers at an address. For each of the three types of census blocks, we ask a straightforward question: Is the available broadband plan, quantified as the average of maximum download speed offered at street addresses in that block for a specific mode of operation (i.e., CAF, monopoly, or competition), better than the other mode(s)? Figures~\ref{fig:q3_pie_1} and~\ref{fig:q3_pie_2} answer this question for Type~A and Type~B census blocks, respectively. We report our findings for download speeds below. We also explored answering this question using the carriage value metric and observed similar trends. 


\smartparagraph{Does CAF ensure better available plans for consumers than monopoly-served non-CAF neighbors?}
In Type~A blocks, where the CAF-supported ISP functions as either a regulated or unregulated monopoly, we anticipated that addresses served through the CAF program would consistently receive superior service compared to those under monopoly. Contrary to expectations, our findings reveal that only about 27~\% of the 8.7~k surveyed census blocks experience better download speeds through CAF. In 55~\% of these blocks, the offered broadband plans in monopoly-served and CAF addresses are identical, and, surprisingly, in roughly 18~\%, CAF plans are inferior to those of monopoly-served neighbors. We hypothesize that this could be because of variation in deployment dates and technologies (among other possibilities) between CAF and non-CAF addresses, but with the data available, we are unable to validate this hypothesis.

Figure~\ref{fig:caf_mono_speeds} shows the distribution of the average maximum download speeds for CAF- and monopoly-served addresses across Type A census blocks where CAF addresses fare better. We observe that for 90~\% of these census blocks, both the CAF and monopoly average maximum speeds are less than 100~Mbps. Additionally, we observe that while the medians are almost equal, at the 80th percentile, CAF speeds are 20~Mbps higher than monopoly speeds. 
To further quantify the improvement of CAF over their monopoly counterparts, where such an improvement exists, we compute, for each census block, the percentage increase of CAF download speeds over monopoly download speeds, as shown in Figure~\ref{fig:caf_mono_percentage}. We observe that for all the Type A census blocks where CAF fares better, the median percentage increase is 75\% and the 80th percentile percentage increase is 400\%. This suggests that for the Type A blocks where CAF has better speeds, the improvement is substantial. 
For completeness, we perform this analysis on the 18\% of Type A blocks where monopoly-served addresses have better speeds; we present the findings in Figures~\ref{fig:q3_mono_1} and~\ref{fig:q3_mono_2} in the Appendix. We observe 
a median percentage increase of 45~\% and an
80$^{th}$ percentile percentage increase of 130~\%. This suggests that where monopoly speeds exceed CAF speeds, the difference is nominal, especially as compared to the difference in those blocks where CAF speeds exceed monopoly speeds.

\smartparagraph{Does CAF ensure broadband plans for consumers are on par with competitively served neighbors?}
In Type~B blocks, where CAF-supported ISPs face competition, we expected CAF-served addresses to be offered plans comparable to those addresses served by multiple providers. This is because the FCC's rate and service standards are intended to reflect the plans available in competitively-served regions. The modal outcome is a tie between a CAF address and a competitive address. However, Figure~\ref{fig:q3_pie_2} also reveals some diversity across outcomes: the plans available at competitively-served addresses were superior to CAF plans in 31~\% of census blocks, and CAF plans were better in 32~\% of census blocks.

Figure~\ref{fig:caf_comp_speeds} depicts the distribution of the average maximum download speeds for CAF and competitively-served addresses across Type B census blocks where CAF addresses have better average download speeds. In these Type~B blocks, we observe trends similar to those in Type A blocks described previously. About 90~\% of these census blocks have CAF and competitively served average speeds less than 100~Mbps, with median speeds at about 10~Mbps. 
We also perform this analysis on the 31~\% of Type B blocks where competitively-served addresses have better speeds, and present the findings in Figures~\ref{fig:q3_comp_1} and~\ref{fig:q3_comp_2} in the Appendix.

\begin{figure}[t]
\centering
\begin{subfigure}[b]{.48\linewidth}
    \includegraphics[width=\linewidth]{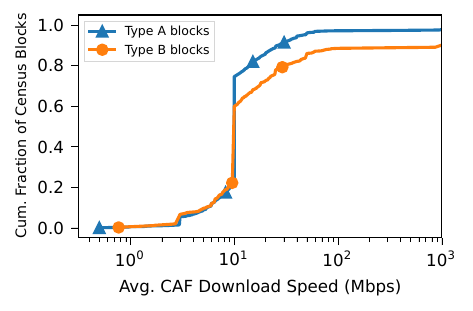}
    \caption{CAF speeds}
    \label{fig:caf_perf_mc}
\end{subfigure}
\begin{subfigure}[b]{.48\linewidth}
    \includegraphics[width=\linewidth]{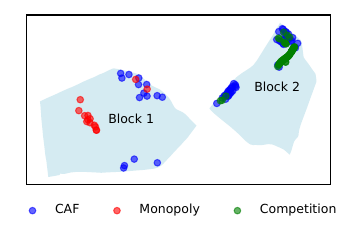}
    \caption{CenturyLink, Georgia.} 
    \label{fig:caf_perf_block}
\end{subfigure}
\caption{Comparison of the performance of CAF addresses across Type A (CAF+Monopoly) and Type B (CAF+Competition) blocks.}
\label{fig:q3_final}
\end{figure}

\smartparagraph{Does CAF offer similar maximum download speeds  in Type~A and B blocks?}
Next, we study whether competition affects service quality offered by CAF-funded ISPs.
In Figure~\ref{fig:caf_perf_mc}, we show the distribution of the average CAF download speeds in Type A and Type B blocks. While there is little difference between the speeds for 60~\% of the blocks, we find that in 20~\% blocks, CAF speeds of Type~B blocks surpass those of Type A blocks by over 90~Mbps. This suggests that CAF addresses in Type~B census blocks experience better speeds than those in Type~A blocks.

To contextualize this finding, we examine two adjacent census blocks, Block~1 (Type~A) and Block~2 (Type~B), served by CenturyLink in Georgia. Notably, competition is present only at the periphery of Block~2, influencing service quality. Proximity to competitively-served addresses correlates with higher download speeds, suggesting that infrastructure investments aimed at outperforming competitors do not extend beyond immediately adjacent addresses. ISPs have minimal incentives to invest in broadband infrastructure where they do not face competition. As a result, the average download speed for CAF-served addresses in Block~2 (100~Mbps) is nearly sixfold that in Block~1 (15~Mbps).

\smartparagraph{Takeaways.}
This analysis demonstrates that CAF-served addresses significantly benefit from nearby competition among service providers. Given that most census blocks currently fall into Type~A, lacking block-level competition, one important implication for policymakers is to foster competition and thereby induce greater ISP infrastructure investment across competitors. Barring the presence of competition, our results suggest that monopoly regulation can help to improve service quality overall, albeit inconsistently.

\vspace*{-0.1in}

\section{Related Work}

\smartparagraph{Internet measurements.}
Previous work  developed BQT and used it to curate a broadband affordability dataset for urban regions in the US~\cite{Paul:Sigcomm23}. In contrast, our study primarily targets US rural regions. Internet availability, quality and  variability have been analyzed by location, demographic variables, and price  in~\cite{feamster_speedtest,Paul:Sigcomm23,ookla_40bn,la_study,paul_speedtest,bauerTPRC,alabamaSpeed}). 
One study~\cite{bat} revealed that the FCC National Broadband Report overestimates internet access and pointed to the lower coverage rates nationwide in 
marginalized communities. Similar discrepancies have been identified in the FCC map for mobile networks~\cite{three_dataset_cacm}. The authors of~\cite{lee2023analysis} analyzed nearly 1~million Ookla Speedtest measurements to explore regional sampling bias and the relationship between internet performance and demographic variables, revealing links between internet speed and demographic factors and emphasizing the need to account for sampling bias. A related study~\cite{fastly:speed} investigated the income-download speed relationship across US zip codes, 
finding a positive correlation between income levels and download speeds. Publicly available data from Ookla was analyzed in~\cite{mapping_Bronzino} analyzed  and  variation in internet quality metrics between communities with different median incomes was identified. Likewise,~\cite{paul_mlab} analyzed M-Lab speed test data in California and observed higher internet quality in urban and affluent areas.


\smartparagraph{CAF program assessment.} 
In the wake of the FCC's announcement launching CAF, some scholars offered hypotheses on the benefits and shortcomings of this transition from the Universal Service Fund's (USF) previous high-cost program to CAF~\cite{ali2020politics, dippon2010replacement, gilroy2012ruralbroadband}. Some scholars have correlated CAF with improvements in rural broadband download speeds~\cite{skorup2020fcc}. Others have used USAC's dataset, treating its data as given, to analyze the relationship between broadband access and wage and employment outcomes~\cite{nazareno2021effectsbroadband}. However,  we have not found a study that attempts to systematically assess the efficacy of CAF by verifying ISP serviceability claims or  compliance with the FCC's rate and service standard. Our approach to our third question, which evaluates the extent to which the FCC's rate and service conditions curb a subsidized monopolist's monopoly powers, builds upon prior work~\cite{narechania2022convergence}. That small-scale study, which relied upon manual queries to ISP websites to develop a dataset of broadband plan information for 126 street addresses across seven states, was only suggestive of broader and more general trends. As noted above, our work here is significantly more robust, querying several orders of magnitude more addresses.


\vspace*{-0.1in}
\section{Implications and Conclusions}
Our analysis sheds light on the effectiveness of the CAF program in its aim to augment broadband availability in underserved communities. Our analysis  compared   data collected from ISP websites to the ISP self-reported data in the USAC dataset. A significant finding is that 44.55~\% of the addresses that are supposed to be served by our studied ISPs in our studied states remain unserved, a fact that contradicts the certifications provided by ISPs to regulators. Additionally, many of the served addresses (i.e. 66.97~\% of the CAF-addresses) receive internet service that falls short of the FCC's minimum speed requirements, highlighting a discrepancy between ISP-reported information to USAC and actual service offered to customers. Our comparison of broadband plans at CAF locations to nearby monopoly-served and competitively-served addresses reveals that competition is most effective at improving consumer value. In census blocks where the CAF-funded ISP operates without competition, the FCC's rate and service conditions can be effective at improving broadband service, but only inconsistently. 

This study is the first to comprehensively evaluate such a large-scale policy intervention. It calls for a re-evaluation of the existing regulatory oversight framework, suggesting a need for more thorough post-hoc verification mechanisms of ISP claims, and advocates for making these audit reports more transparent to the public and policymakers. Indeed, federal and state officials should consider past compliance with funding programs such as CAF when deciding how to allocate new funds.  Moreover, the FCC should consider revising its rate and service quality requirements to ensure that regulated monopolies provide better value to the underserved communities targeted by this program, and policymakers should consider ways to foster greater competition in monopoly-served regions.

Our findings also have broader implications beyond CAF. They demonstrate the importance of exhaustive post-hoc evaluation mechanisms, which could offer meaningful regulatory oversight for other policy interventions, particularly those focused on availability and affordability. Specifically, the post-hoc evaluation framework developed in this paper could be readily applied to the BEAD program~\cite{ntia_bead_program}, which is 
poised to spend over \$~42 billion. 

In conclusion, our study emphasizes the necessity for rigorous oversight and empirical evaluation in broadband infrastructure policymaking. With substantial public investment in broadband availability, affordability, and adoption, it is vital to assess the impact of these investments and refine strategies for improved outcomes. This requires a commitment to data-driven policy development and the flexibility to adapt based on empirical evidence.

\vspace*{-0.1in}
\section{Acknowledgments}

This work was funded through the National Science Foundation Internet Measurement Research award \#2220417. The Bright Initiative provided access to proxy infrastructure for data collection.

\bibliographystyle{ACM-Reference-Format}
\bibliography{refs}

\vspace{0.1in}
Appendices are supporting material that has not
been peer-reviewed.
\section{Appendix}

\subsection{Limitations}
\label{sec:limitations}

\smartparagraph{Advertised vs. experienced service quality.} Our data only captures the information ISPs advertise to consumers, and it does not always reflect the experienced service quality. Previous work~\cite{bst} has pointed out the gaps between advertised and experienced service quality. Thus, our findings might offer only an optimistic estimate of service quality for CAF users. Given the challenges of reliably measuring service quality at scale, we leave the exploration of bridging the gap between advertised and experienced service quality for future exploration. 

\smartparagraph{Excluded data points.}
Our methodology is specifically designed to conclusively identify addresses not served by an ISP, reporting an address as unserved only when the ISP's website explicitly indicates service unavailability. For instance, Appendix Figure~\ref{fig:bspeed_no_service} illustrates the notification shown on the website for an unserved address. Nonetheless, several of our queries do not offer affirmative answers. Sometimes they fail for unknown reasons and sometimes offer ambiguous responses. For example, we noted that certain ISPs, particularly AT\&T, sometimes redirect our queries to a ``Call to Order'' page. While technically these addresses might be serviceable within the FCC's required 10-day window, confirmation would necessitate manual calls to the ISPs, potentially requiring impersonation of residents at these addresses. Due to scalability challenges and, more critically, ethical concerns, we have opted to exclude these addresses from our analysis, instead resampling another street address from the same census block group. Consequently, our serviceability reporting is subject to errors, depending on whether the ISPs can meet the FCC's 10-day service provision requirement for these addresses. 


\begin{figure}[t]
   \centering
   \includegraphics[width=0.75\linewidth]{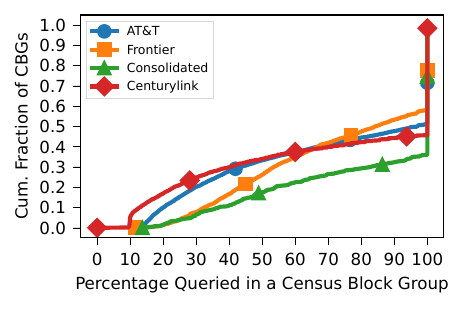}
   \caption{CDF of the percentage of addresses queried for each ISP.}
   \label{fig:CDF_Quer}
\end{figure}

\smartparagraph{Aggregation at finer spatial granularity.}
Reporting various metrics of interest, such as the carriage value, maximum download speed, serviceability rate, etc. at any spatial granularity requires computing various aggregate statistics, such as mean, median, max, etc. To ensure statistical significance for these results, similar to the methodology in~\cite{Paul:Sigcomm23}, our goal is to collect the maximum between 30 or 10~\% of the addresses in a CBG. However, in some instances we are unable to capture that number of samples; several factors, such as fewer CAF addresses in a CBG (see Figure~\ref{1c}), lower hit rates (see Section~\ref{ssec:bqt}), etc., contribute to this challenge. This negatively affects the statistical significance of our results in some CBGs in Sections~\ref{sec:q1} and~\ref{sec:q2}. One approach to address this problem could be to consider coarser spatial granularities for aggregation; however, that approach would mix different geographic, demographic, and socioeconomic attributes,  making random sampling of street addresses within a CBG ineffective. We leave the exploration of a more efficient sampling strategy that strikes a balance between the statistical significance of aggregated metrics while preserving the geographic, demographic, and socioeconomic homogeneity in reported statistics for future work. 

\smartparagraph{Veracity of advertised plans.}
There is no system or database available to verify the accuracy of download speed and price data provided by ISPs for specific street addresses. However, as noted in~\cite{bat}, ISPs have little incentive to advertise exaggerated or misleading information about their services to potential customers. 

\smartparagraph{Staleness.}
We began our data collection in June 2023, querying each street address only once. Therefore, our dataset represents a single snapshot of offered plans, some of which may be outdated. It is possible that available plans at some addresses could have changed since our query, leading to a staleness issue. However, 
according to FCC requirements, ISPs that receive funding through the CAF~II model were obligated to meet prescribed service quality requirements by December 2021. Since our querying commenced well after this deadline, our reports of non-compliance, while in some cases not current, are representative.

\begin{figure}[t]
   \centering
   \includegraphics[width=0.75\linewidth]{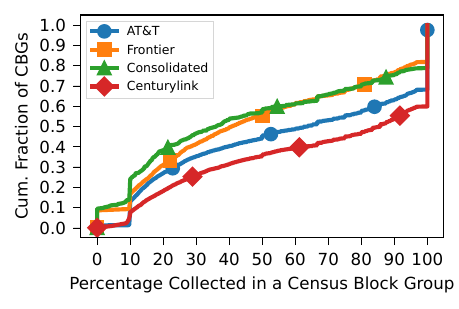}
   \caption{CDF of percentage of addresses collected for each ISP.}
   \label{fig:CDF_Coll}
\end{figure}

\begin{table*}[t]
  \scriptsize
  \centering
  \caption{Errors in Traceback.}
  \label{tab:tracebacks}
  \begin{tabular}{c|c|c|c|c|c}
     \multicolumn{1}{c}{}   & \multicolumn{1}{c}{\textbf{Select Drop-down Address}} & \multicolumn{1}{c}{\textbf{Analyzing Result}} & \multicolumn{1}{c}{\textbf{Empty traceback}}& \multicolumn{1}{c}{\textbf{Clicking Button}}  & \multicolumn{1}{c}{\textbf{Other Error}}  \\
     
    \hline
    AT\&T (61,768) & 43,781 & 10,130 & 7,606 &-& 14  \\
    Frontier (26,791) & 17,614 & - & 6,210 &2,967& - \\
    CenturyLink (6,939) & - & - & 6,939 & - & -\\
    Consolidated (15,551) & 15,510 & 33 & - & - & 8 \\
   
    \hline
  \end{tabular}
\end{table*}

\smartparagraph{Unavoidable issues during data collection}
While we attempted to query addresses in accordance with the sampling strategy outlined in Section~\ref{ssec:sampling}, several challenges and errors prevented us from doing so completely in all geographies. In particular, we encountered unavoidable errors for 8,164 Frontier addresses, and a significant proportion of these addresses were concentrated in census block groups in Wisconsin. For these CAF addresses, we found that the dropdown box did not appear when the address was typed in the search bar. We verified this error both by using BQT and manually, and for Frontier in particular, re-queried this entire set of addresses at least two times to verify that the error persisted. Furthermore, these addresses could not be resampled as they were often in either small census block groups or almost all of the replacement addresses had the same issue. 
Additionally, for AT\&T, Consolidated, and CenturyLink, we were unable to collect data from 2,304, 5,191, and 1,758 addresses respectively. Nevertheless, as Figure~\ref{fig:CDF_Quer} indicates, we were able to query enough addresses to meet our sampling goals for AT\&T and Consolidated. For CenturyLink, we were unable to query 10~\% of addresses in 215 census block groups due to issues with human verification challenges. Figure~\ref{fig:CDF_Coll} illustrates the results of these queries, after filtering out those queried addresses that returned repeated errors during data collection. It showcases that for AT\&T and CenturyLink, we were able to collect 10\% of the addresses per census block group for almost all census block groups. In comparison, for Frontier and Consolidated less than 10\% of the addresses were collected in approximately 10\% of the census block groups. 


\smartparagraph{Breakdown of errors experienced during data collection.}
As noted above and elaborated in the Appendix, many addresses were classified as unknown due to errors during data collection. In table~\ref{tab:tracebacks}, we categorize these unknown addresses by the errors we received in their tracebacks, i.e. the text we return after each address query in case an error is encountered. Table~\ref{tab:tracebacks} indicates that the largest proportion of the errors for AT\&T, Frontier, and Consolidated occurred when attempting to select an address from the ISP's drop-down box. More specifically, we found that, in many cases, the address did not exist in the box, so we were unable to proceed to the next page. In the case of AT\&T, we found that a maximum of 10,130 addresses could potentially be attributed to the ''Call to Order'' case that we described earlier.  We believe that this number is small compared to the overall number of addresses we have received data for in AT\&T (176k), and is unlikely to significantly affect the final results. Overall, the errors identified in table~\ref{tab:tracebacks} describe why we were unable to collect the maximum between 30 addresses and 10\% of the addresses in each census block group (in those instances where that proved true).

\begin{figure}[H]

   \includegraphics[width=.75\linewidth]{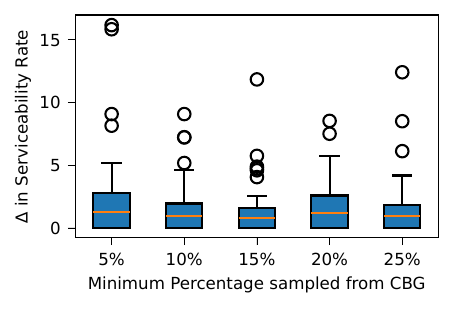}
    \caption{ \(\Delta \) in AT\&T's Serviceability Rate after sampling different percentages from each census block group with more than 30 addresses.}
    \label{fig:sensitivity}
\end{figure}
\begin{figure}[t!]
\centering
\begin{subfigure}[b]{.75\linewidth}
    \includegraphics[width=.96\linewidth]{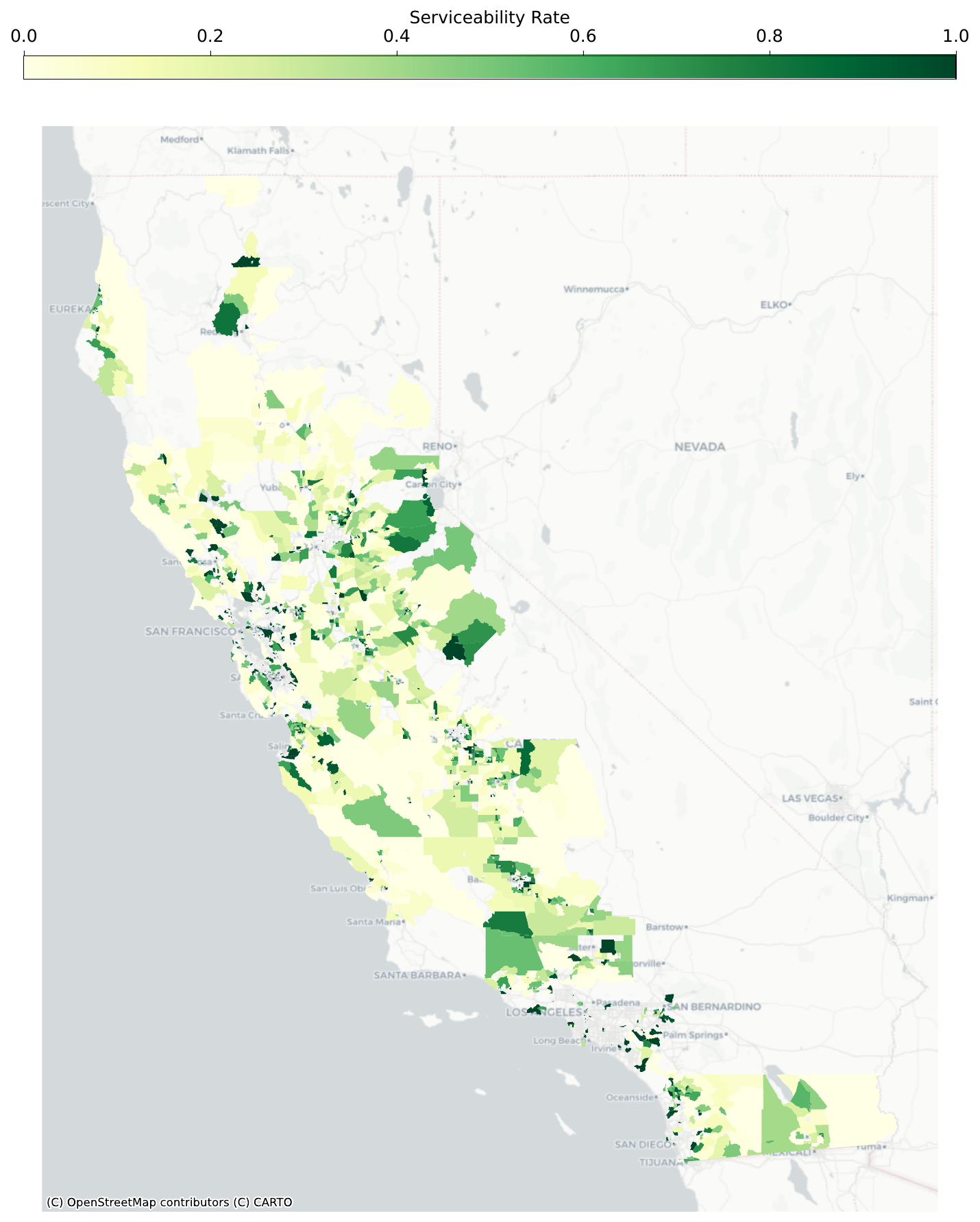}
    \caption{California}
    \label{fig:heatmap_ca}
\end{subfigure}
\begin{subfigure}[b]{.75\linewidth}
   \includegraphics[width=0.96\linewidth]{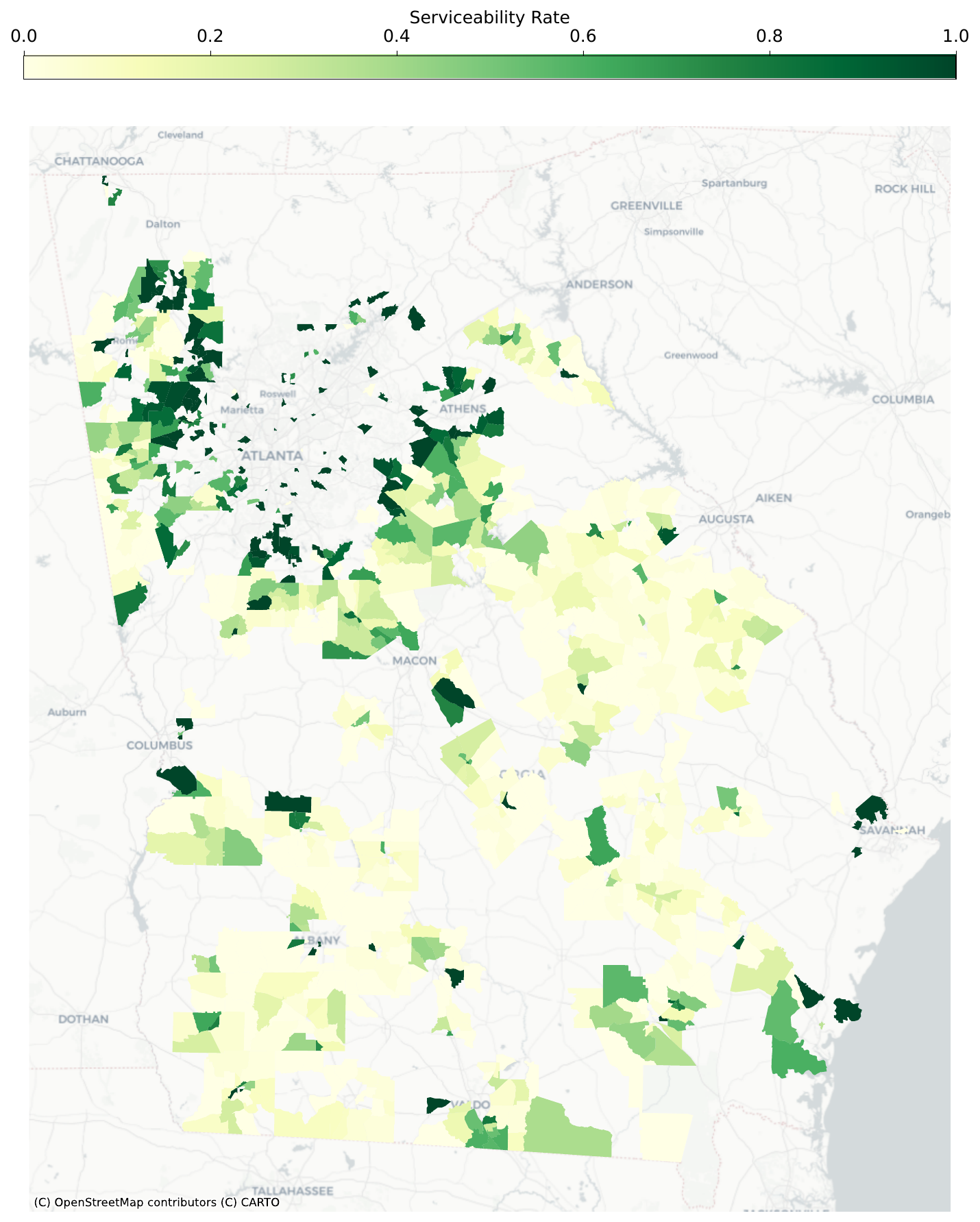}
    \caption{Georgia.}
    \label{fig:heatmap_ga}
\end{subfigure}
\caption{Geospatial distribution of AT\&T's serviceability rates.}
\label{fig:q1_heatmap}
 \vspace*{-0.23in}
\end{figure}
\subsection{Sensitivity Analysis}
\label{ssec:sensitivity}
Given the significance of our findings, we investigate the sensitivity of our results to our sampling strategy. To this end, we randomly select a set of {46 census block groups} (each with more than 30 addresses) and query at least 75\% of the total CAF addresses in each. We then treat the serviceability rates estimated from these samples as ground truth and report the error, quantified as the difference in serviceability rates, for various random sampling strategies, each varying in sampling rates.  We show this result in Figure~\ref{fig:sensitivity}. We observe that errors are less than 5\% for all sampling rates, indicating that our results are robust to changes in sampling rates. This finding also highlights the diminishing returns of querying additional CAF addresses within a census block group.
\begin{figure*}[t]
\centering

\begin{subfigure}[b]{0.24\linewidth}
\includegraphics[width=\textwidth]{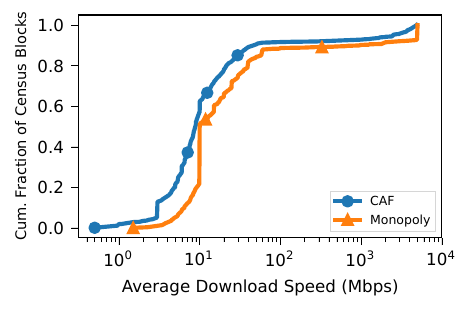}
 \caption{CDF of CAF and monopoly speeds in Type A blocks} 
\label{fig:q3_mono_1} 
\end{subfigure}%
\begin{subfigure}[b]{0.24\linewidth}
\includegraphics[width=\textwidth]{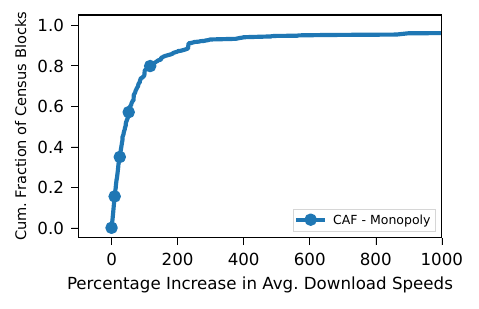}
 \caption{CDF of percentage increase in Type A block speeds} 
\label{fig:q3_mono_2} 
\end{subfigure}%
\begin{subfigure}[b]{0.24\linewidth}
\includegraphics[width=\textwidth]{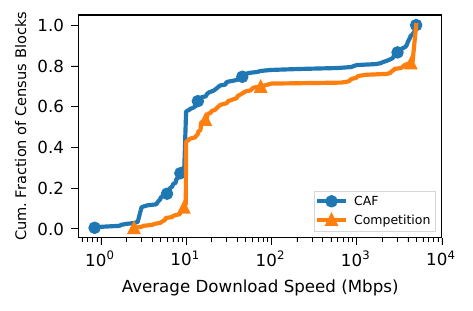}
\caption{CDF of CAF and competition speeds in Type B blocks}
\label{fig:q3_comp_1}
\end{subfigure}%
\begin{subfigure}[b]{0.24\linewidth}
\includegraphics[width=\textwidth]{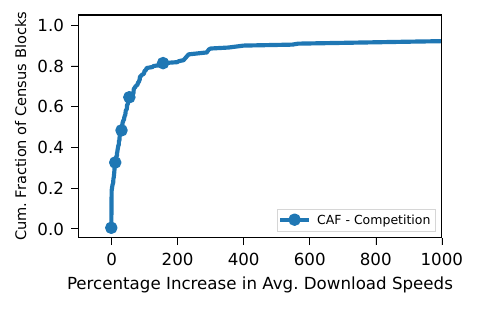}
\caption{CDF of percentage increase in Type A block speeds}
\label{fig:q3_comp_2}
\end{subfigure}
\caption{Distribution of the average maximum speeds and percentage increase of speeds in Type A and Type B blocks where CAF performs worse than its counterpart.
\label{fig:q3_comp_mono}}
\end{figure*}

\begin{table*}[t]
  \scriptsize
  \caption{Total number of CAF addresses collected per ISP per state.}
  \label{tab:data}
  \resizebox{\textwidth}{!}{%
  \begin{tabular}{l|c c c|c c c|c c c|c c c}
    \multicolumn{1}{c}{} 
    & \multicolumn{3}{c}{\textbf{AT\&T}} 
    & \multicolumn{3}{c}{\textbf{CenturyLink}} 
    & \multicolumn{3}{c}{\textbf{Consolidated}} 
    & \multicolumn{3}{c}{\textbf{Frontier}} \\
    & \textbf{Street Addresses} & \textbf{BGs} & \textbf{CBGs} 
    & \textbf{Street Addresses} & \textbf{BGs} & \textbf{CBGs} 
    & \textbf{Street Addresses} & \textbf{BGs} & \textbf{CBGs} 
    & \textbf{Street Addresses} & \textbf{BGs} & \textbf{CBGs} \\
    \hline
    California & 69,711 & 10,707 & 1,759 & - & - & - & - & - & - & 48,447 & 8,786 & 664 \\
    Georgia & 37,772 & 6,344 & 753 & 464 & 74 & 19 & - & - & - & 850 & 82 & 14 \\
    Illinois & 8,745& 2,124 & 303 &  1,461 & 478 & 66 & 1,332 & 480 & 39 & 33,260 & 8,394 & 681 \\
     New Hampshire & - & - & - & - & - & - & 7,229 & 1,154 & 175 & - & - & - \\
    North Carolina &  12,525 & 1,153 & 215 & 28,411 & 3,623 & 812 & - & - & - & 7,834 & 591 & 106 \\
    Ohio & 22,185 & 3,711 & 542 & 25,780 & 5,083 & 639 & - & - & - & 49,631 & 6,665 & 558 \\
    Utah & - & - & - & 1,749 & 498 & 178 & - & - & - & 2,332 & 531 & 28 \\
    Alabama & 23,862 & 4,869 & 669 & 10,083 & 3,211 & 427 & 295 & 57 & 5 & 4,401 & 670 & 56 \\
    Florida & 11,029 & 1,829 & 344 & 10,104 & 2,845 & 625 & 4,010 & 535 & 49 & 578 & 136 & 5 \\
    Iowa & - & - & - & 9,757 & 3,700 & 624 & - & - & - &  4,092 & 1,720 & 89 \\
    Mississippi & 38,069 & 9,208 & 950 & 2 & 1 & 1 & - & - & - & 1,237 & 197 & 20 \\
    Nebraska & -& - & - & 3,986 & 1,666 & 261 & - & - & - & 2,648 & 1,208 & 63 \\
    New Jersey & - & - & - & 980 & 269 & 88 & - & - & - & - & -& - \\
    Vermont &-& - & - & - & - & - & 9,940 & 1,502 & 201 & - & - & - \\
    Wisconsin & 9,349 & 2,287 & 303 & 19,064 & 7,850 & 686 & - & - & - & 14,456 & 2,621 & 224 \\
    \hline
    \textbf{Total} & 233,247 & 42,232 & 5,838 & 111,841 & 29,298 & 4,426 & 22,806 & 3,728 & 469 & 169,766 & 31,601 & 2,508\\
  \end{tabular}}
\end{table*}

\begin{table*}[t]
  \scriptsize
  \centering
  \caption{Total number of addresses queried for Question 3.}
  \label{tab:q3}
  \begin{tabular}{l|cc|cc|cc|cc|cc|cc}
     \multicolumn{1}{c}{}   & \multicolumn{2}{c}{\textbf{AT\&T}} & \multicolumn{2}{c}{\textbf{Frontier}} & \multicolumn{2}{c}{\textbf{CenturyLink}} & \multicolumn{2}{c}{\textbf{Consolidated}} & \multicolumn{2}{c|}{\textbf{Xfinity}} & \multicolumn{2}{c}
    {\textbf{Spectrum}} \\
     & {\textbf{CAF}} & \textbf{Non-CAF} &
     {\textbf{CAF}} & \textbf{Non-CAF} & 
     {\textbf{CAF}} & \textbf{Non-CAF} & 
     {\textbf{CAF}} & \textbf{Non-CAF} &
     {\textbf{CAF}} & \textbf{Non-CAF} & 
     {\textbf{CAF}} & \textbf{Non-CAF} 
     \\ \hline
    \hline
    California & 39,894 & 22,071 & 30,360 & 8,843 & - & 211 & - & 57 & - & 9,608 & - & 6,096 \\
    Georgia & 20,303 & 12,034 & 494 & 444 & 306 & 675 & - & 7 & - & 2,158 & - & 1,066 \\
    Illinois & 2,824 & 1,452 & 14,345 & 6,988 &  373 & 422 & 406 & 137 & - & 163 & - & 249 \\
    North Carolina & 8,716 & 5,530 & 3,878 & 3,045 & 21,757 & 22,341 & - & - & - & 186 & - & 7,067 \\
    New Hampshire & - & - & - & - & - & - & 2,665 & 1,570 & - & 112 & - & 447 \\
    Ohio & 13,852 & 4,691 & 36,710 & 16,206 & 18,356 & 7,553 & -  & 892 & - & 503 & - & 5,673 \\
    Utah & - & - & 741 & 193 & 603 & 517 & - & - & - & 573 & - & - \\
    \hline
    \textbf{Total} & 85,589 & 45,778 & 86,528 & 35,719 & 41,395 & 31,719 & 3,071 & 2,663 & - & 13,303 & - & 20,598 \\
    \hline
  \end{tabular}
\end{table*}

\subsection{Data Collection Workflow}
\label{ssec:datacollectionworkflow}
\begin{figure}[b]
   \centering
\includegraphics[width=0.33\textwidth, height=0.20\textheight]{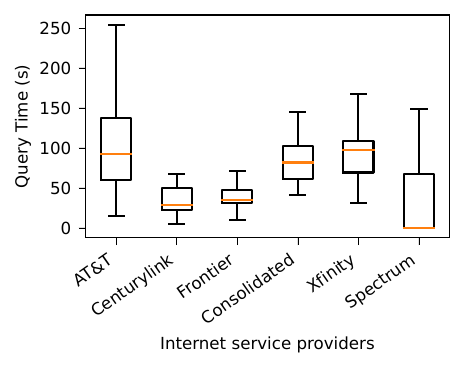}
    \caption{Per-address query times for each ISP.} 
    \label{fig:query_times}
\end{figure}

\smartparagraph{Broadband-plan querying tool (BQT).}
Our approach leverages and augments the broadband-plan querying tool (BQT)~\cite{Paul:Sigcomm23}  to obtain broadband plan information (i.e., speed and price data) for a set of ISPs at any US street address.  As described in~\cite{Paul:Sigcomm23}, BQT takes as input a street-level address and returns the set of broadband plans (upload/download speeds and corresponding prices in US dollars) offered by an ISP at that address. Through automated mimicking of the behavior of a real user interacting with the ISP's website, BQT can ease the aggregation of available plan information at scale. We would also like to highlight through Figure~\ref{fig:query_times} that the specific workflow for each ISP leads to different query times for the queried addresses. We found that AT\&T in particular had a wide distribution of query times due to the anti-bot detection mechanisms that were implemented. In the following, we describe the specific query process for each ISP in our study, the set of possible responses, and the action we take in each case.





\begin{figure*}[t]
\begin{subfigure}[t]{0.4\linewidth}
    \includegraphics[width=1\linewidth]{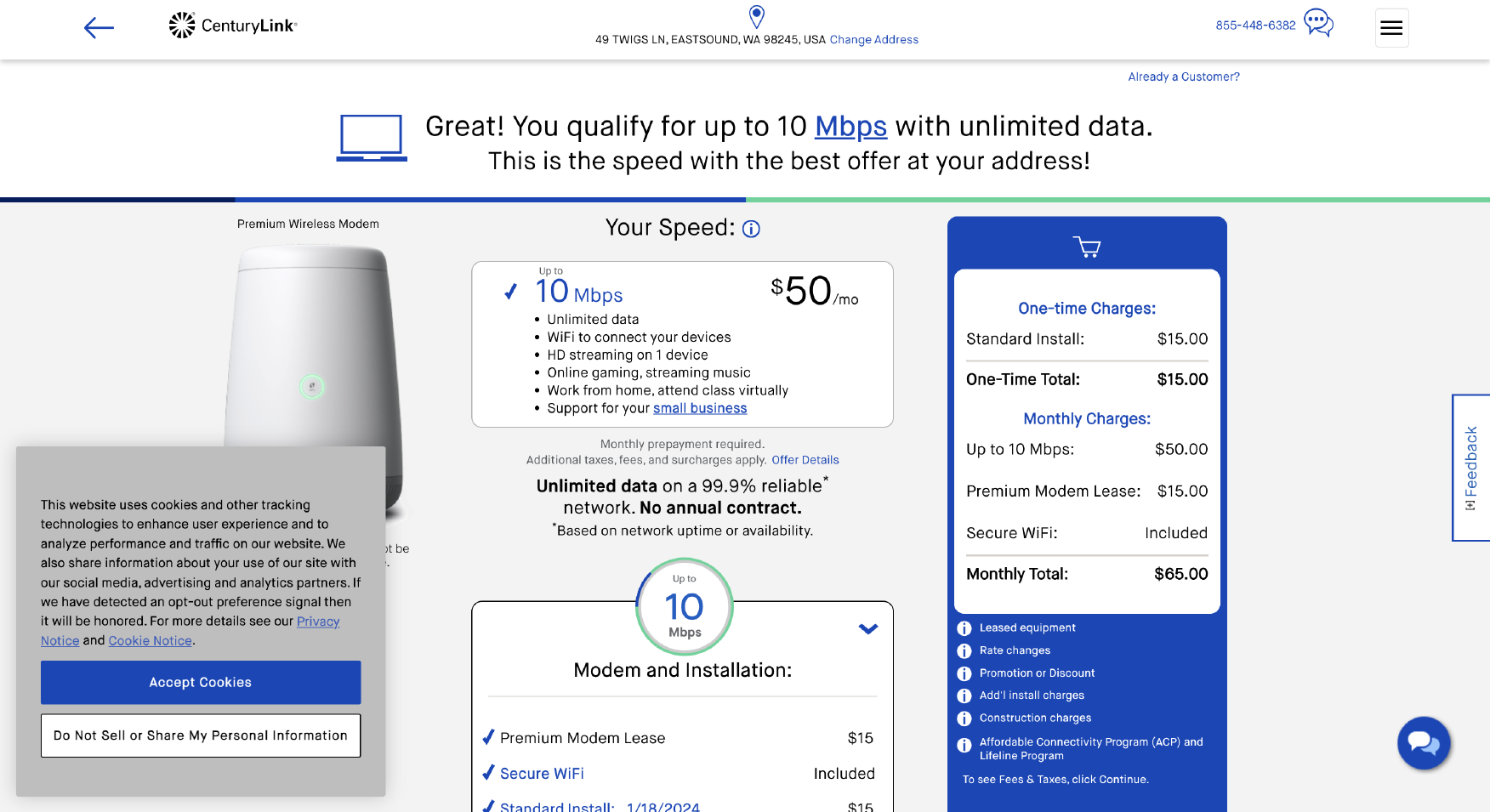}
    \caption{CenturyLink webpage displaying the available plans for the specified address.}
    \label{fig:clink_served}
\end{subfigure}
\hspace{0.8in}
\begin{subfigure}[b]{0.4\linewidth}
    \includegraphics[width=1\linewidth]{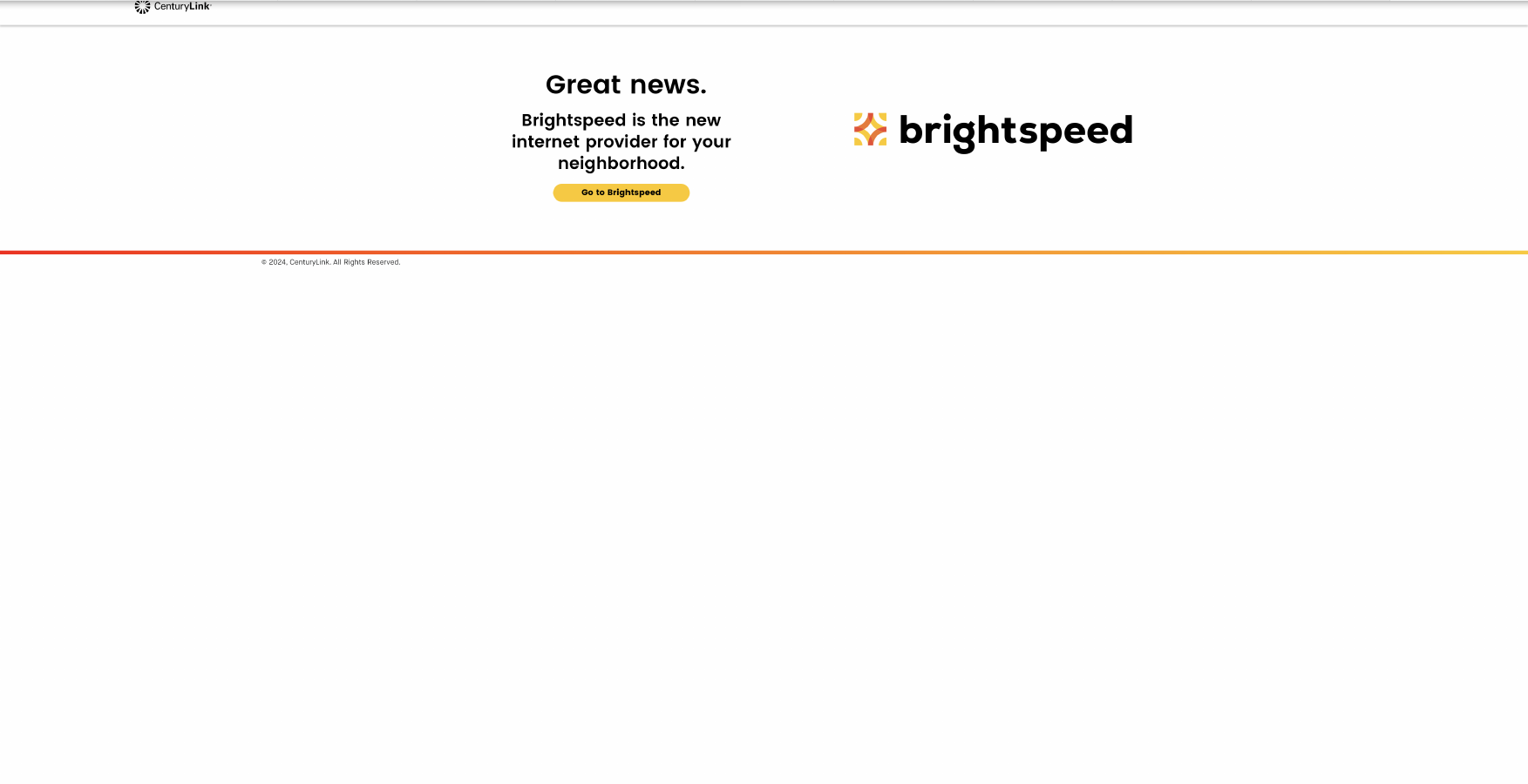}
    \caption{Webpage redirecting CenturyLink users to Brightspeed.}
    \label{fig:clink_redirect}
\end{subfigure}
\hspace{0.8in}
\begin{subfigure}[b]{0.4\linewidth}
    \includegraphics[width=1\linewidth]{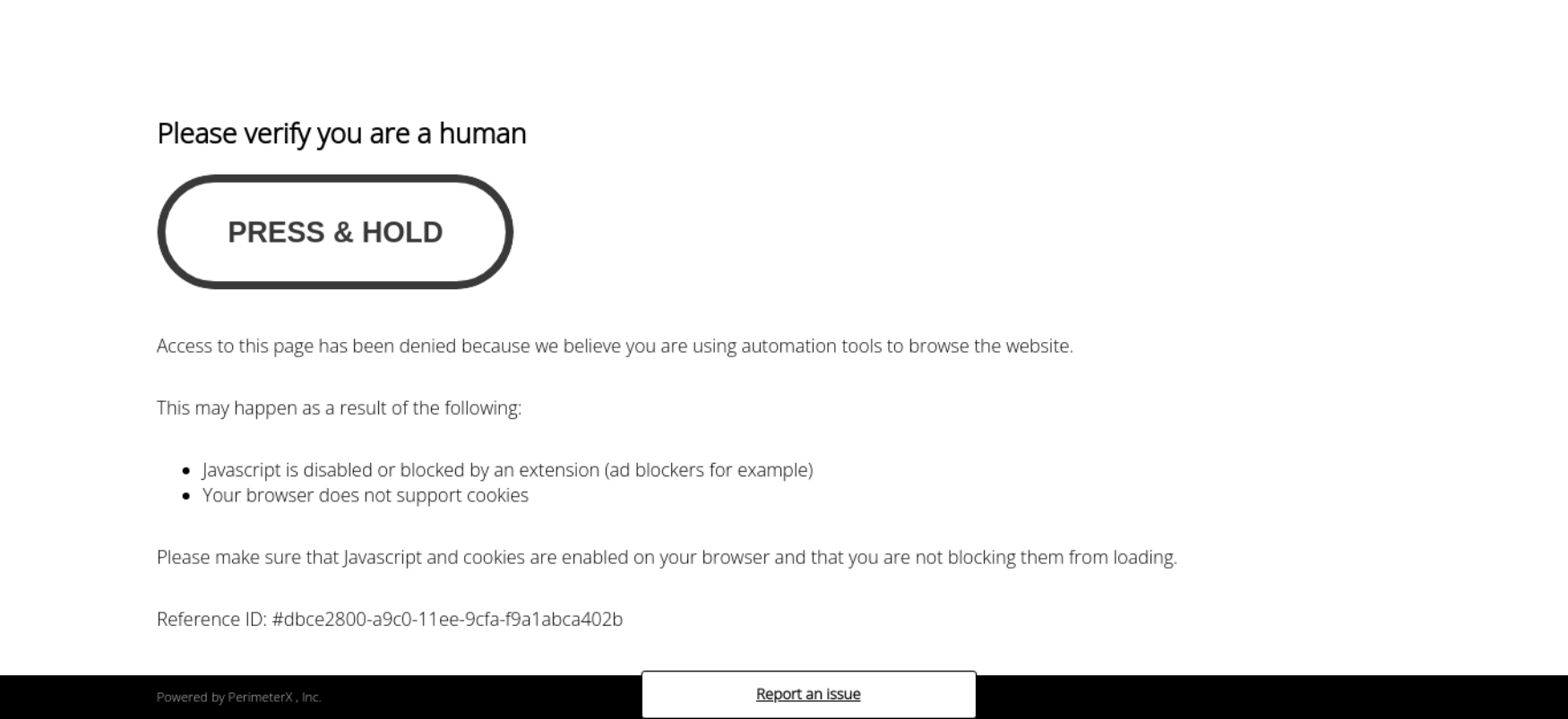}
    \caption{CenturyLink's page for ``human verification''.}
    \label{fig:clink_verification}
\end{subfigure}
\hspace{0.8in}
\begin{subfigure}[b]{0.4\linewidth}
    \includegraphics[width=1\linewidth]{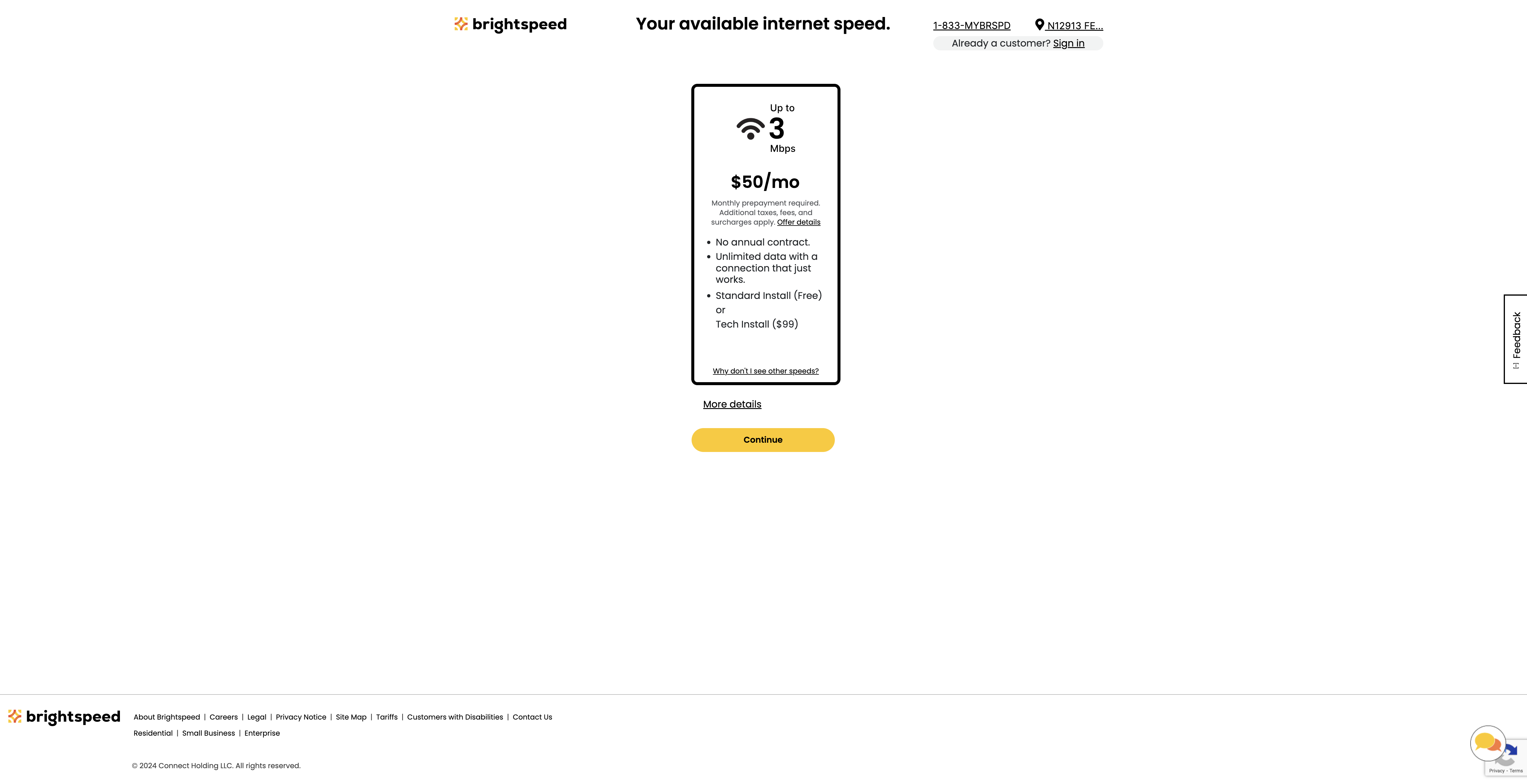}
    \caption{Brightspeed webpage displaying plans for an address served by CenturyLink. }
    \label{fig:bspeed_serviced}
\end{subfigure}
\hspace{0.8in}
\begin{subfigure}[b]{0.4\linewidth}
\centering
    \includegraphics[width=1\linewidth]{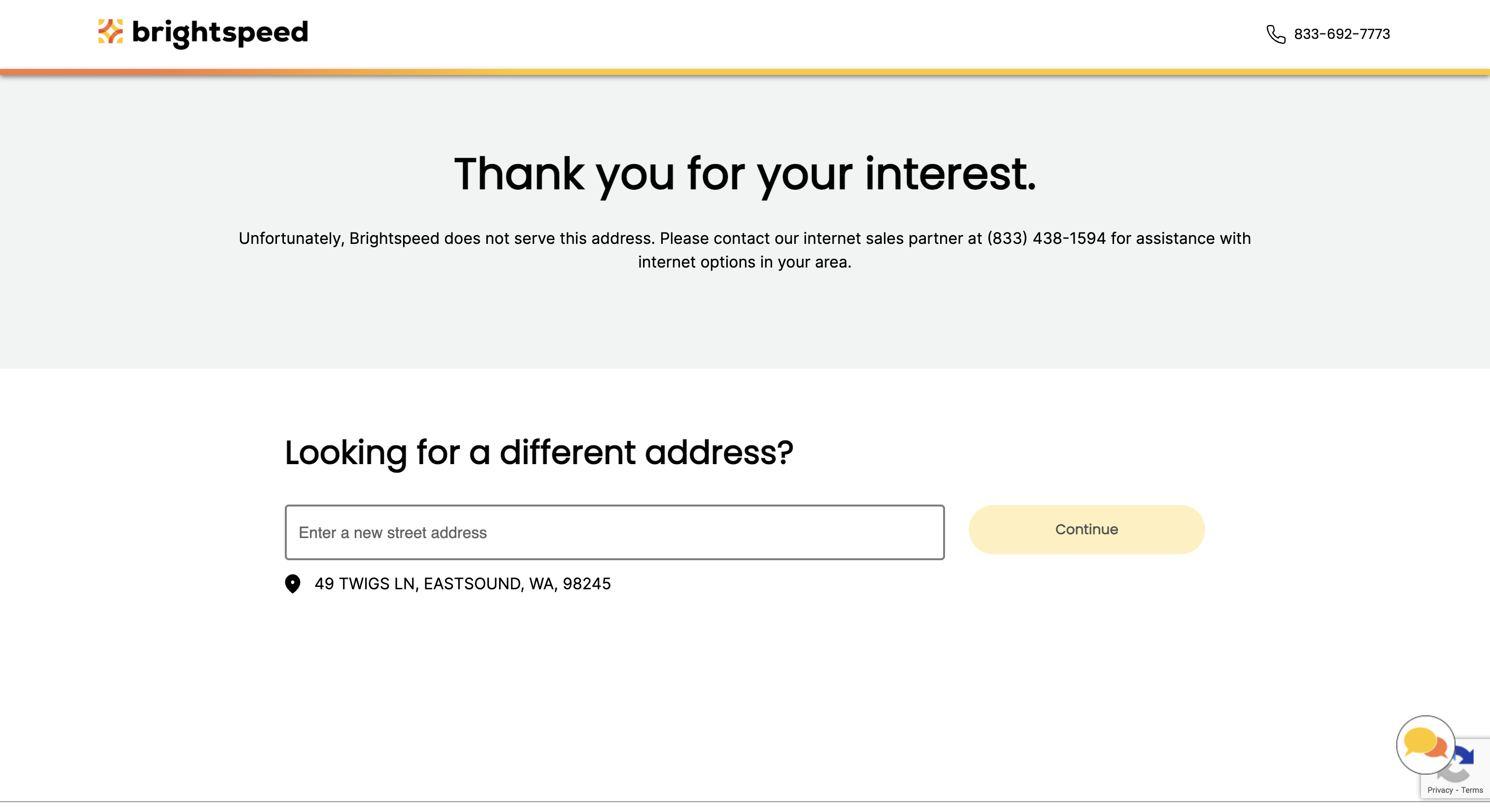}
    \caption{Brightspeed webpage indicating that the specified address is unserviced.}
    \label{fig:bspeed_no_service}
\end{subfigure}
\caption{Example webpages from the CenturyLink and Brightspeed querying processes.}
\label{fig:clink_bspeed_querying}
\end{figure*}

\smartparagraph{CenturyLink:} 
CenturyLink provided service to CAF addresses in 12 of the 15 states in our study. Because CenturyLink sold some of its assets and obligations, including those associated with the CAF program, to Brightspeed (a different ISP brand), we determined the service availability and compliance for CenturyLink by querying both the ISP websites. \footnote{The ISPs CenturyLink and Brightspeed have the same parent company, Lumen Technogies}Specifically, we began by querying a CAF address on the CenturyLink website. If the address was serviced by CenturyLink we were redirected to a webpage displaying the plans, as seen in Figure~\ref{fig:clink_served}, and we logged  the address  as serviceable. In many instances, however, CenturyLink's website directed users to Brightspeed's website instead, as shown in Figure~\ref{fig:clink_redirect}. In this case, we then queried Brightspeed's website with that address. One roadblock we encountered throughout this process was a ``human verification'' page, as seen in Figure~\ref{fig:clink_verification}. Because the tool was then unable to proceed further, this was logged as an error. 

After we queried all the addresses with CenturyLink, the addresses not logged as ``Serviceable" were subsequently queried on Brightspeed's website. In one case, Brightspeed's website returned a webpage that displayed the broadband service plans available at that address, as shown in  Figure~\ref{fig:bspeed_serviced}. Here we logged the address as ``Serviceable" and collected data about available broadband service plans. Finally, if neither CenturyLink's nor Brightspeed's website indicated that broadband service was available at the queried address, then the address was logged as having ``No Service" as seen in Figure~\ref{fig:bspeed_no_service}. 

\begin{figure*}[p]
\centering
\begin{subfigure}[b]{0.4\linewidth}
   \includegraphics[width=\textwidth]{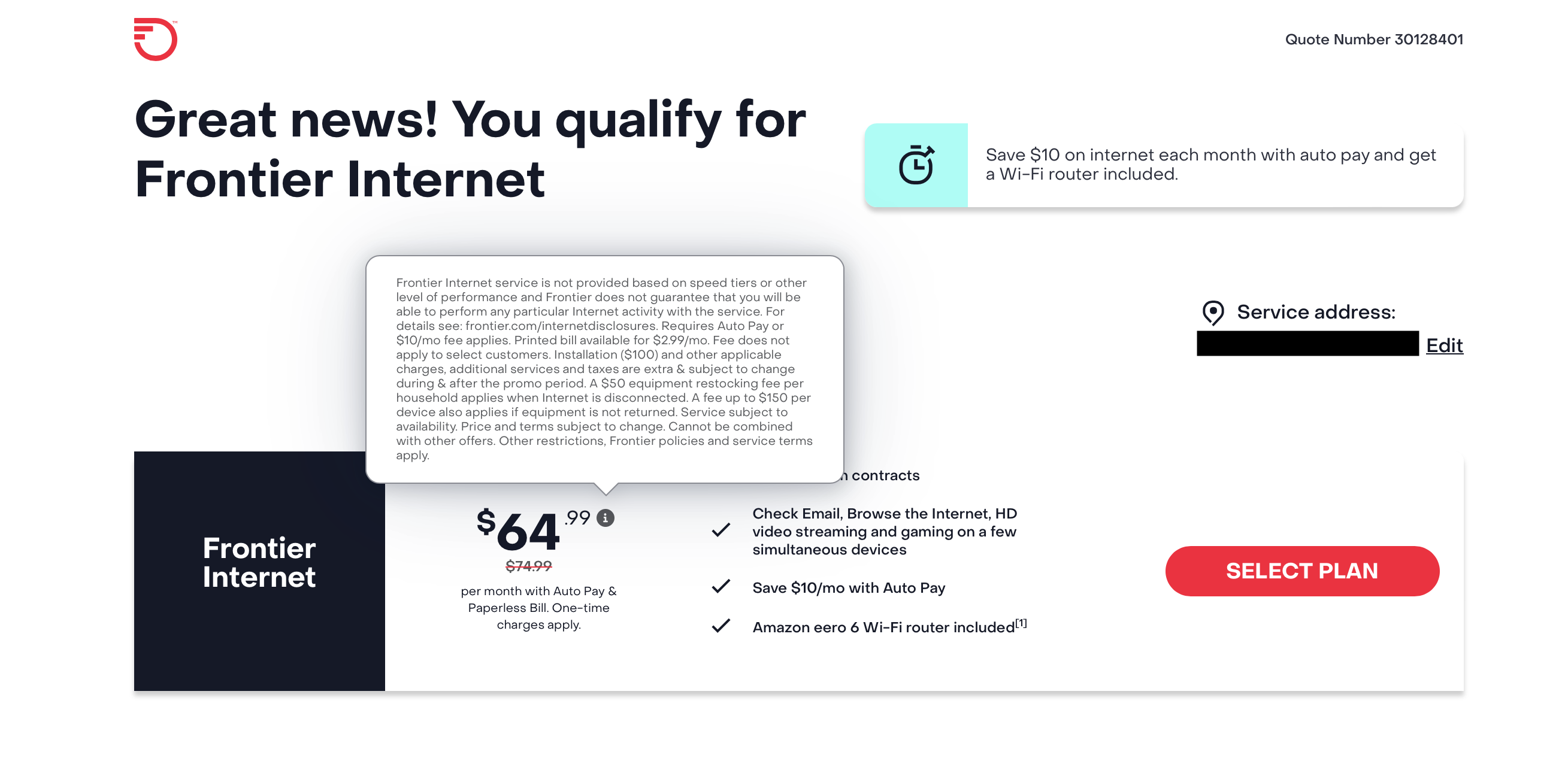}
    \caption{Webpage displaying a note about ``Frontier Internet."} 
    \label{fig:frontier_plan_information} 
\end{subfigure}%
\hspace{0.8in}
\begin{subfigure}[b]{0.4\linewidth}
     \includegraphics[width=\textwidth]{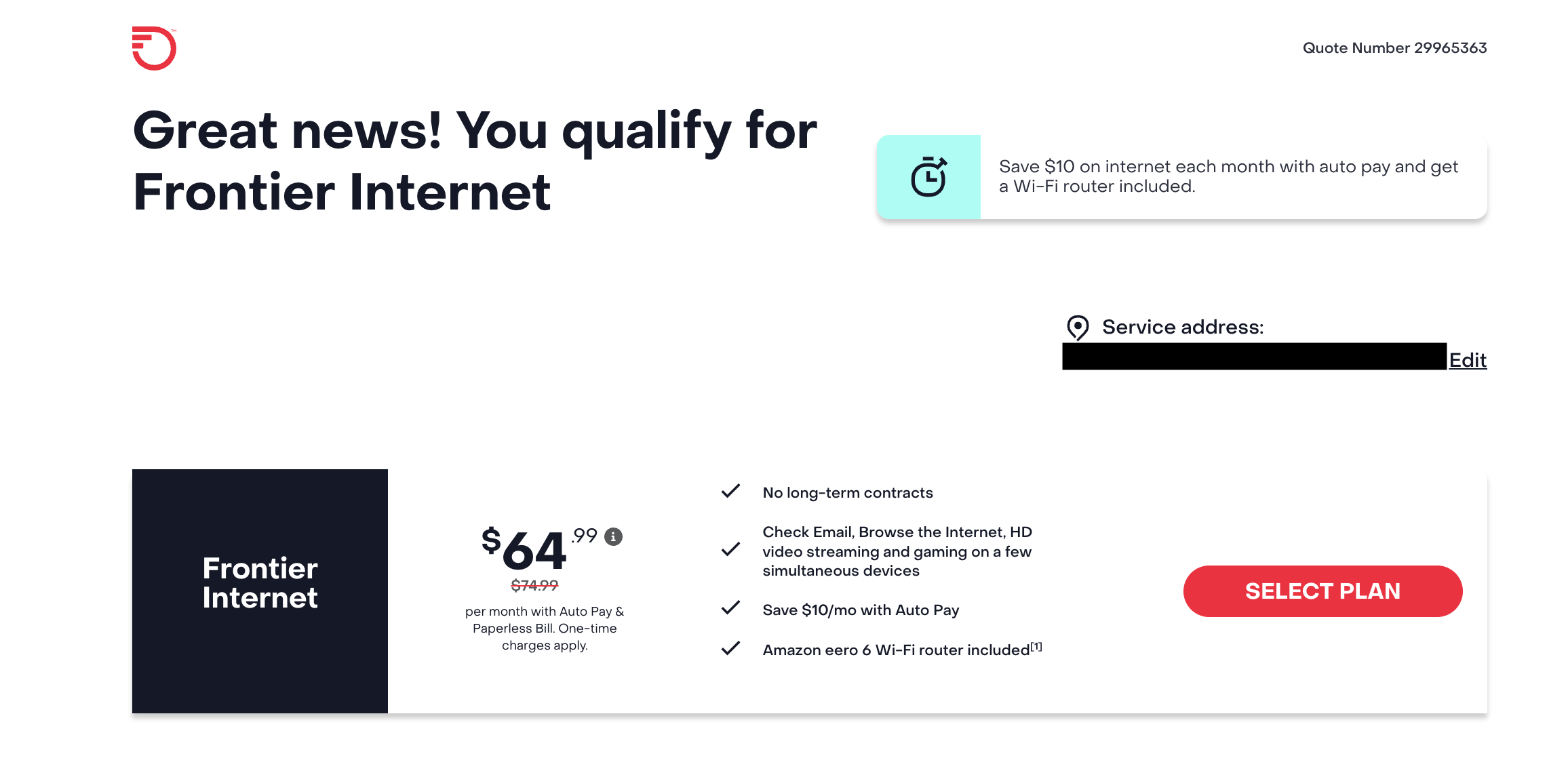}
    \caption{Webpage displaying the available plans for the specified address.} 
    \label{fig:frontier_serviced} 
\end{subfigure}

\medskip

\begin{subfigure}[b]{0.4\linewidth}
    \includegraphics[width=\textwidth]{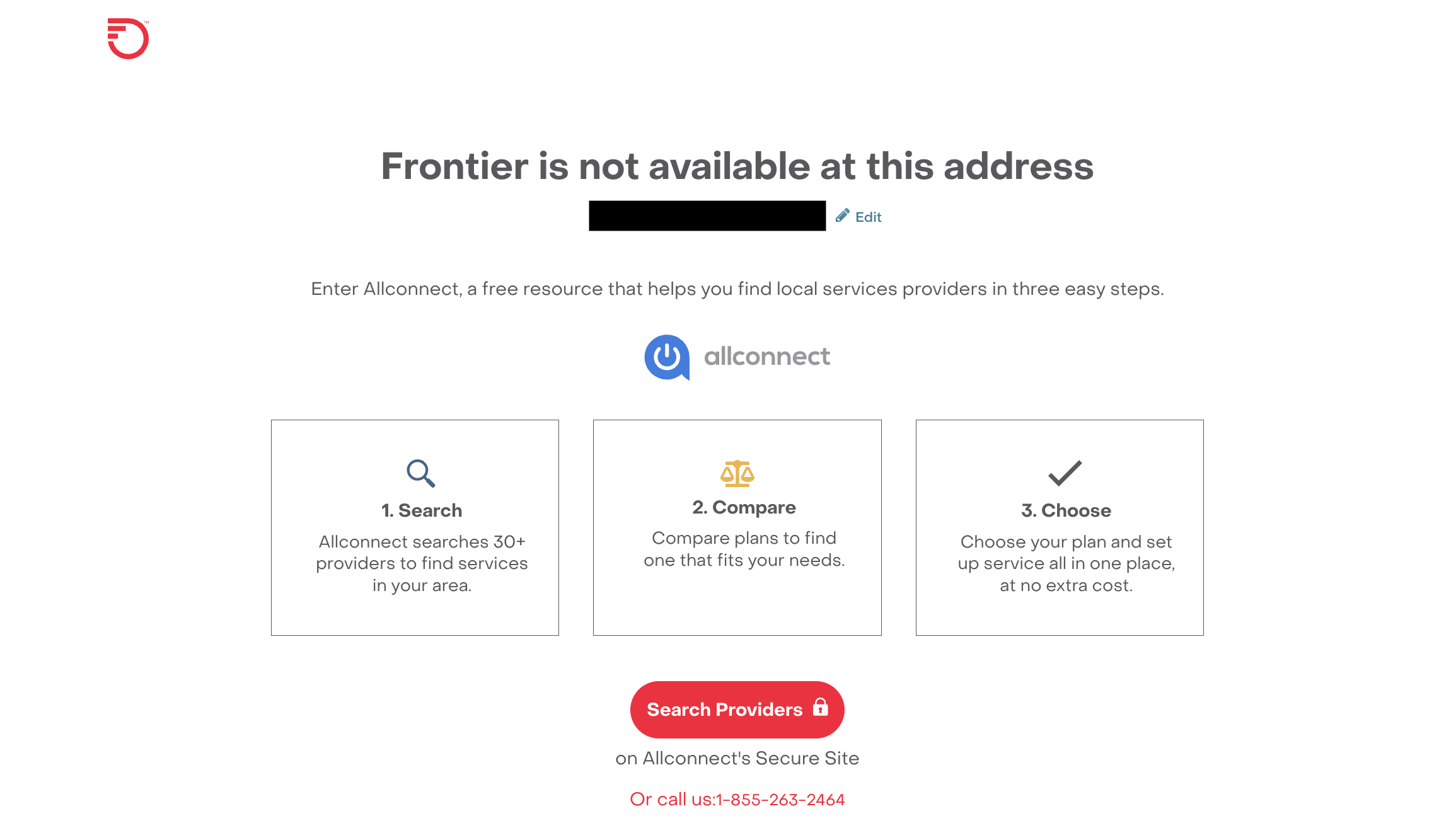}
    \caption{Webpage indicating an address is unserviced.}
    \label{fig:frontier_no_service}
\end{subfigure}%
\hspace{0.8in}
\begin{subfigure}[b]{0.4\linewidth}
    \includegraphics[width=\textwidth]{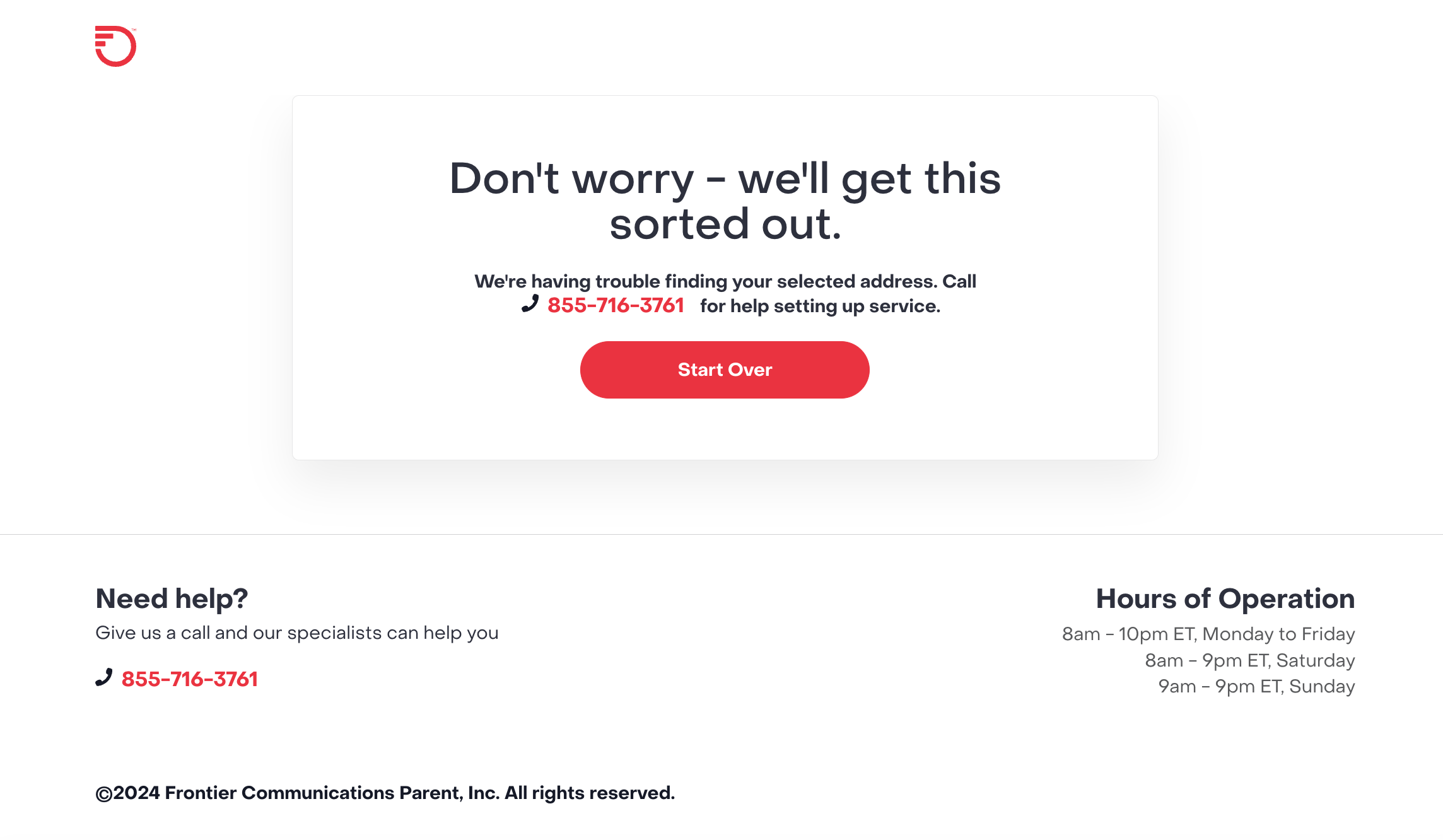}
    \caption{Webpage indicating an address is not found in the BQT dropdown menu.}
    \label{fig:frontier_address_not_found}
\end{subfigure}

\vspace{-5pt}
\caption{Example webpages from the Frontier querying process.}
\label{fig:Frontier_querying}
\end{figure*}

\smartparagraph{Frontier:}
Frontier offered plans varying from 500 Mbps to 5000 Mbps. These plans were offered to CAF addresses in 12 of the 15 states in our study. Some of these plans were designated as ``Frontier Internet'', which, as described in Section~\ref{sec:q2}, did not offer any guarantee to consumers of a minimum download speed. An example is shown in Figure~\ref{fig:frontier_plan_information}. We note that Frontier has previously settled cases alleging that it fraudulently over-promised and under-delivered on the download and upload speeds its networks were capable of offering to consumers~\cite{frontiercommunications2022}.

The BQT queries to Frontier's website
returned three possible results, which we logged as ``Serviceable'', ``No Service'', and ``Address Not Found''. If Frontier's website returned a webpage displaying broadband service plans available at that address, as shown in  Figure \ref{fig:frontier_serviced}, we logged the address as ``Serviceable" and collected data about available broadband service plans, including whether the plan offered was the ``Frontier Internet'' plan, or whether it was another plan guaranteeing a minimum download speed. If Frontier's website returned a webpage that indicated that no broadband plans were available at that address, as shown in  Figure \ref{fig:frontier_no_service}, we logged the address as  ``No Service''. Finally, in some instances, Frontier's website could not resolve the address as entered through BQT (Frontier's website used a dynamic dropdown menu to resolve addresses), shown in  Figure~\ref{fig:frontier_address_not_found}. In this instance, the address was logged as ``Unknown''.

\begin{figure*}[p]
\centering
\begin{subfigure}[b]{0.4\linewidth}
    \includegraphics[width=1\linewidth]{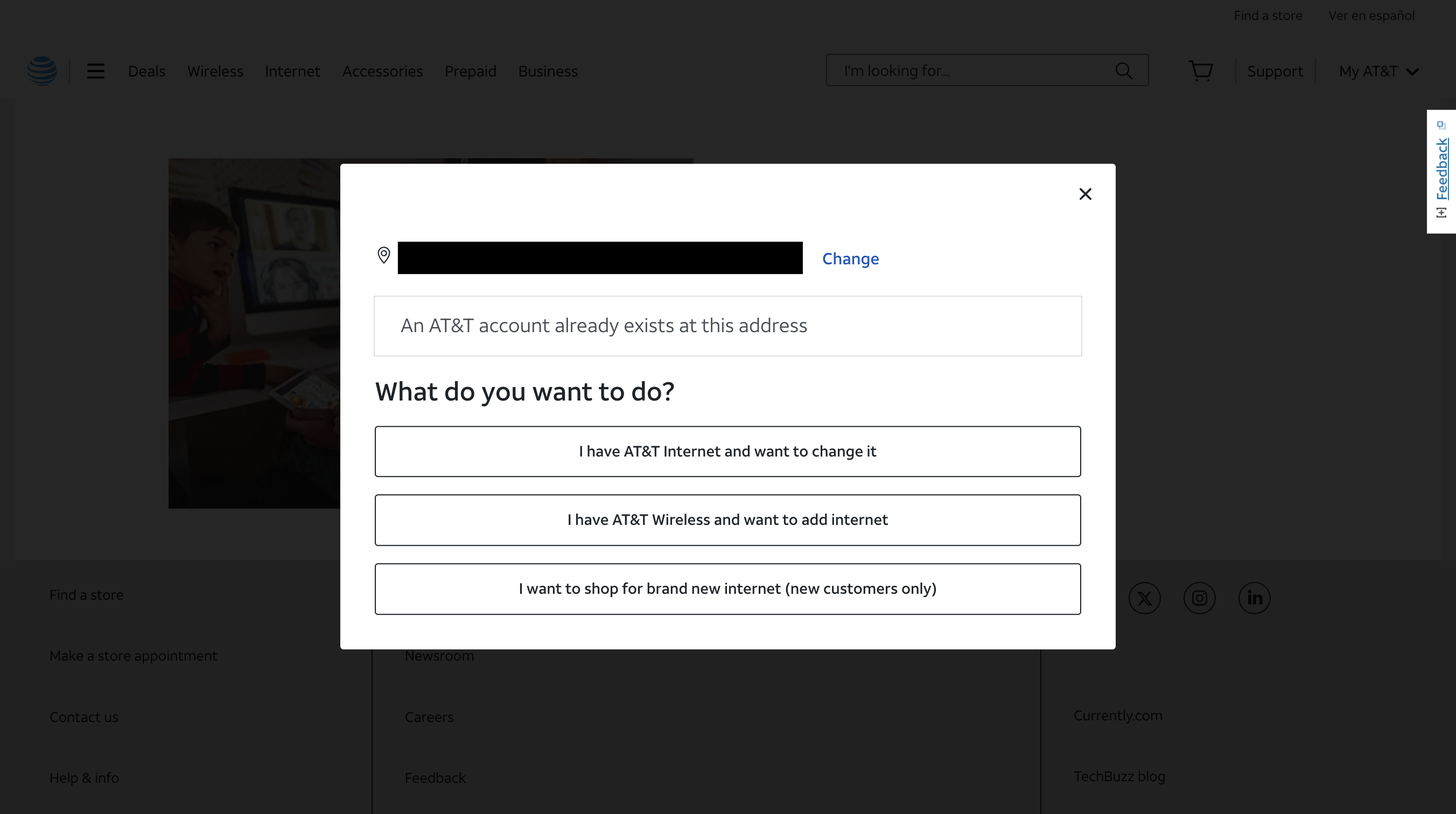}
    \caption{ Webpage displaying an existing plan at the specified address.\label{fig:att_existing}}
\end{subfigure}
\hspace{0.8in}
\begin{subfigure}[b]{0.4\linewidth}
    \includegraphics[width=1\linewidth]{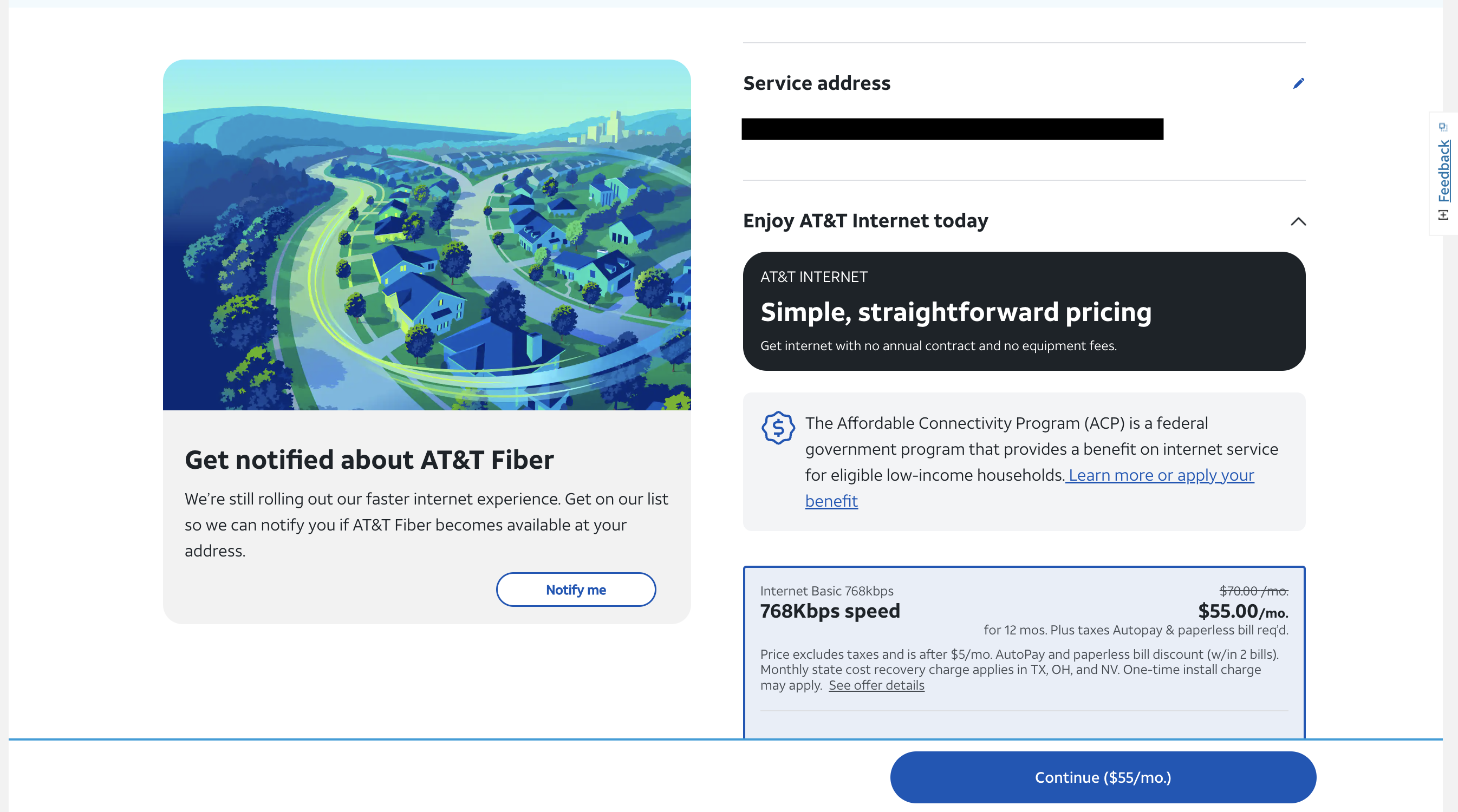}
    \caption{ Webpage displaying the available plans for the specified address.\label{fig:att_plans}}
\end{subfigure}%
\hspace{0.8in}
\begin{subfigure}[b]{0.4\linewidth}
    \includegraphics[width=1\linewidth]{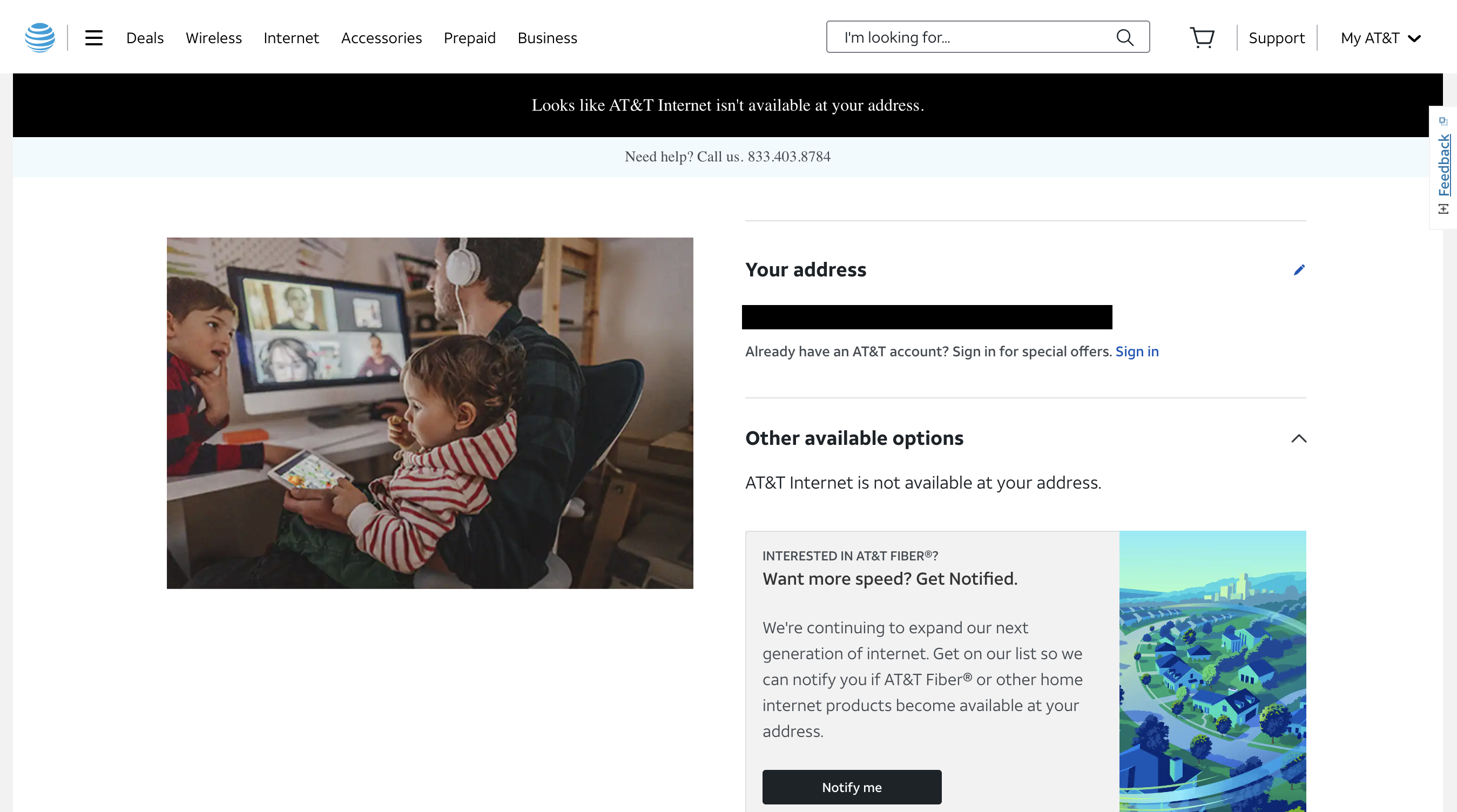}
    \caption{ Webpage indicating that no plans are available at the specified address.\label{fig:att_no_service}}
\end{subfigure}%
\hspace{0.8in}
\begin{subfigure}[b]{0.4\linewidth}
    \includegraphics[width=1\linewidth]{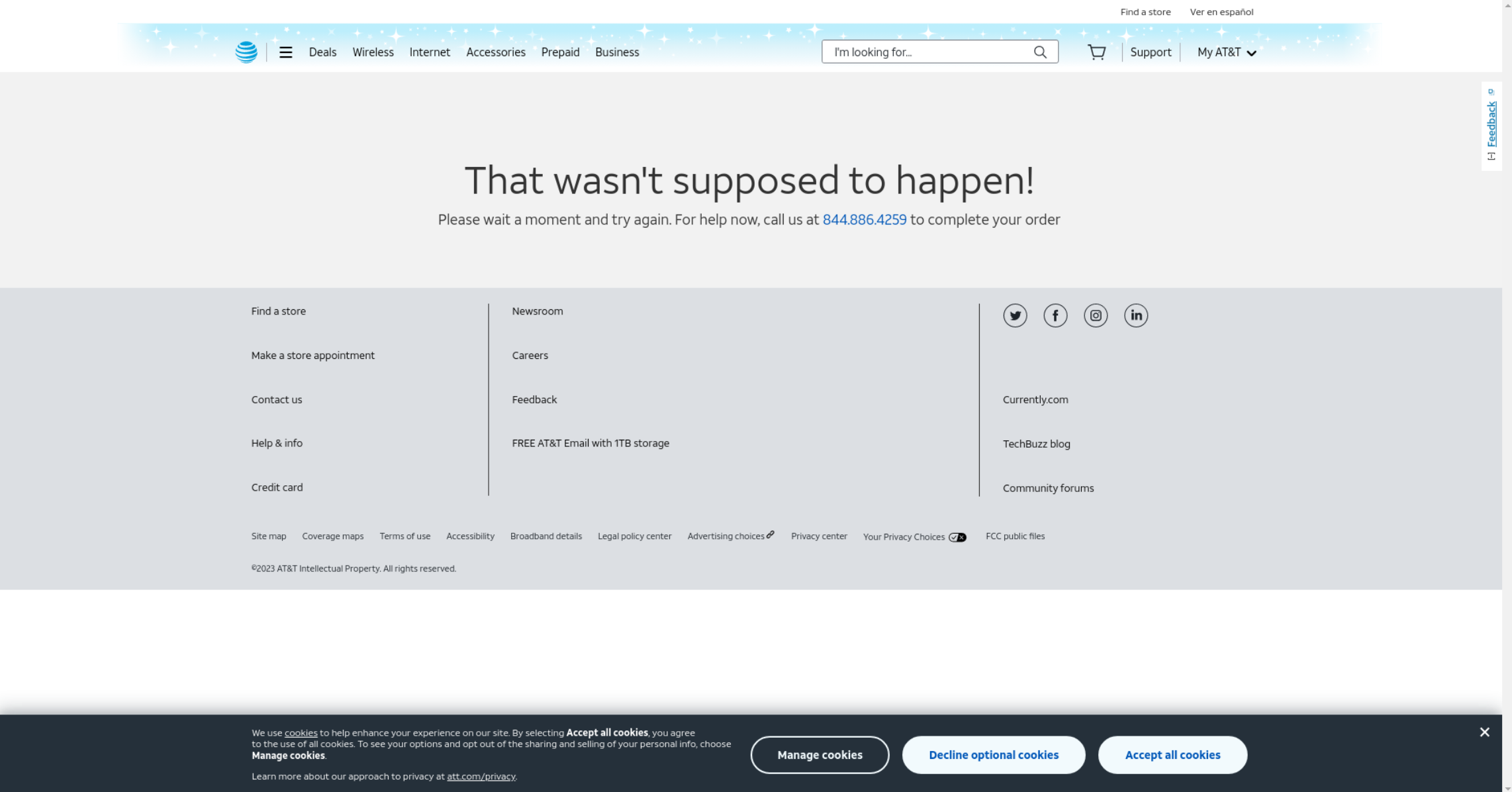}
    \caption{ ``Call to Order'' webpage that requires the user to call to obtain internet plan information.\label{fig:att_call}}
\end{subfigure}
\caption{Example webpages from the AT\&T querying process.}
\vspace{-5pt}
\end{figure*}

\smartparagraph{AT\&T:}
AT\&T offered a range of plans to CAF addresses in 9 of the 15 states in our study, with download speeds varying from 768~Kbps to 5~Gbps, and includes an ``Internet Air'' plan that, as described in Section~\ref{sec:q2}, does not appear to offer consumers any guaranteed minimum speed. Like with the other ISPs, we used BQT to query AT\&T's website to determine service availability and compliance. BQT entered each address into a web form on AT\&T's website and then attempted to select that address from the form, which AT\&T's website dynamically resolved in a dropdown element. BQT either selected the address from the dropdown or, if the address was not in the dropdown, it attempted to proceed with the complete address as entered to view available plans. AT\&T's website then returned one from among a range of possible responses. In some cases, AT\&T already had an active subscriber at the queried address and asked whether the user would like to modify the service or look for a new plan, as pictured in  Figure~\ref{fig:att_existing}. In this case, BQT selected the option to look for a new plan in order to see all available options for that address. We then logged the address as ``Serviceable" and collected plan information. In cases where there was no active subscriber, AT\&T’s website may return a webpage displaying broadband service options available at that address, as shown in Figure~\ref{fig:att_plans}. In such cases, we logged the address as ``Serviceable," and collected data about those plans. In other such cases, AT\&T's website indicated that service is not available at the queried address, as shown in Figure~\ref{fig:att_no_service}, in which case we logged the address as ``No Service''. Finally, AT\&T's website occasionally returned an ambiguous result. For example, an ambiguous response, depicted in  Figure~\ref{fig:att_call}, directed website users to call AT\&T and speak with a representative to get information about service plans available at that address. We logged responses such as these as ``Unknown."

\begin{figure*}
\begin{subfigure}[b] {0.4\linewidth}
    \includegraphics[width=.8\linewidth]{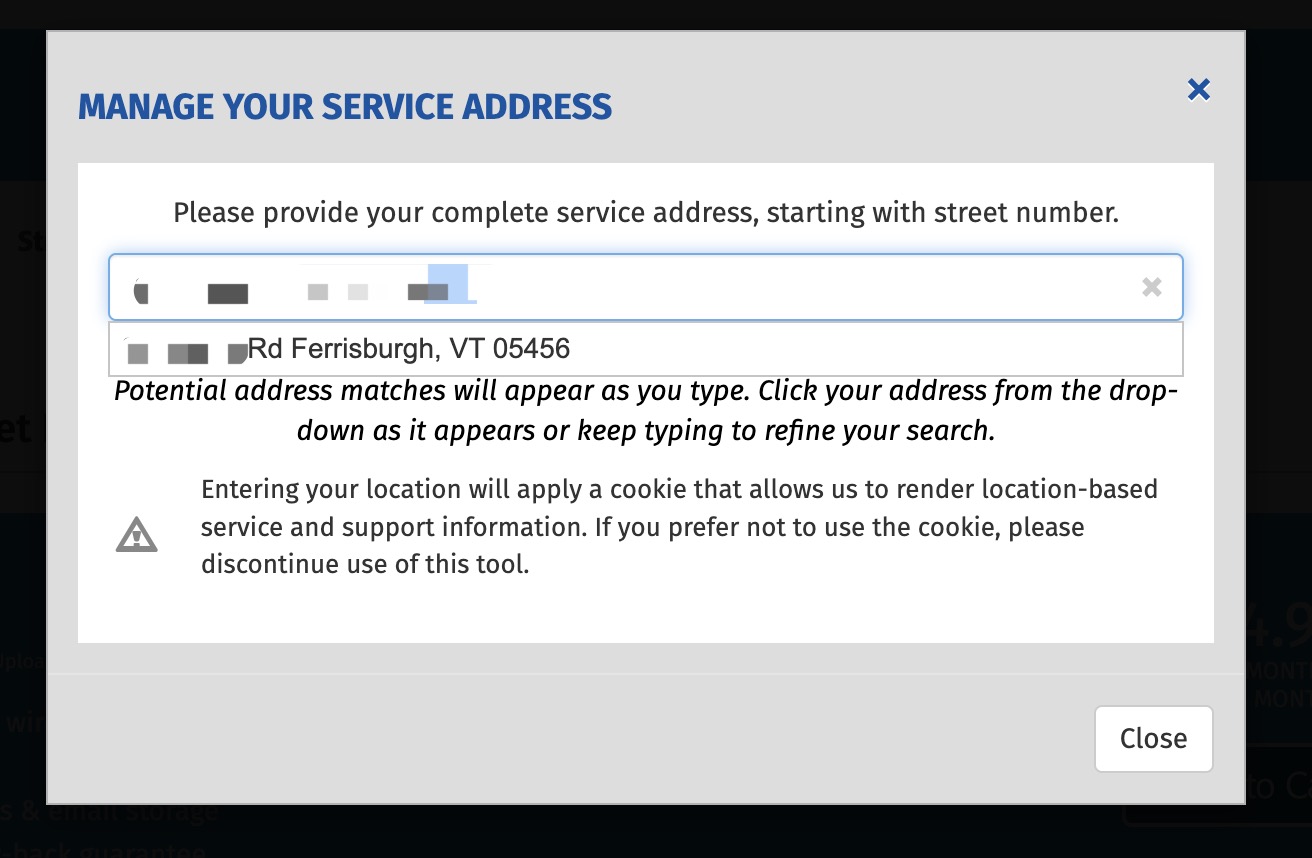}
    \caption{Address suggestion format.}
    \label{fig:Consolidated Communication Address Suggestion Format}
\end{subfigure}
\hspace{0.8in}
\begin{subfigure}[b] {0.4\linewidth}
    \includegraphics[width=.9\linewidth]{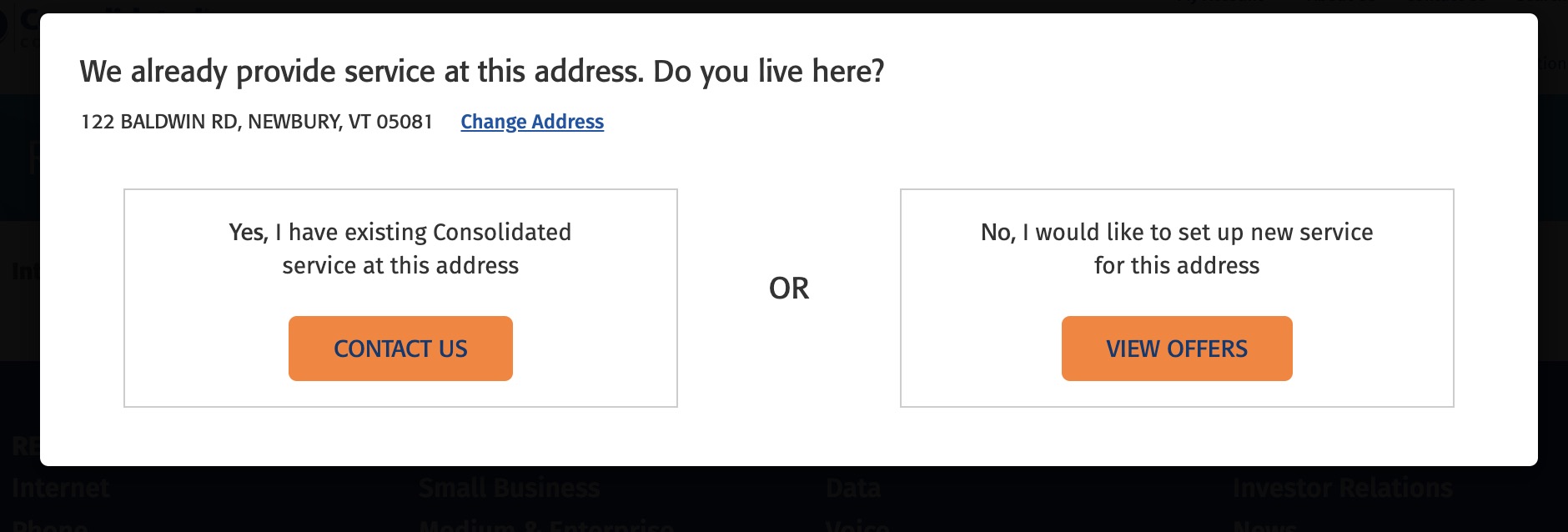}
    \caption{Indicating existing user.}
    \label{fig:Consolidated Communication Already Provide Service Modal}
\end{subfigure}
\hspace{0.8in}
\begin{subfigure}[b] {0.4\linewidth}
    \includegraphics[width=.9\linewidth]{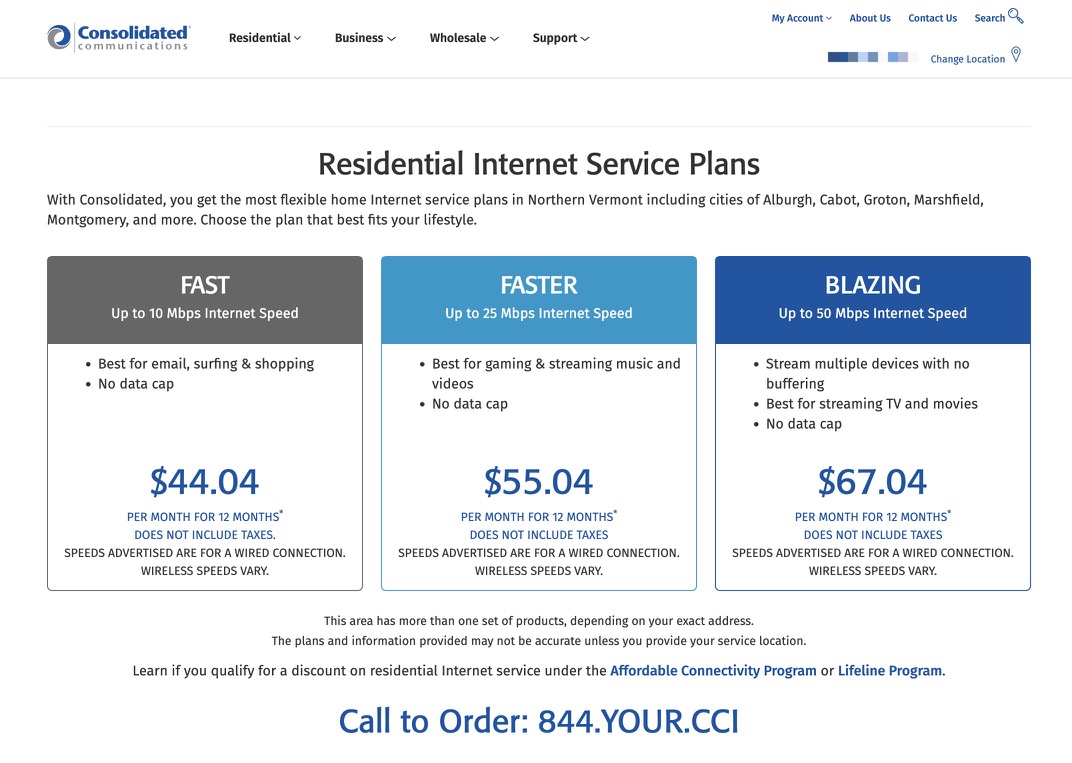}
    \caption{Displaying available plans.}
    \label{fig:Consolidated Communication New Plan}
\end{subfigure}
\hspace{0.8in}
\begin{subfigure}[b] {0.4\linewidth}
    \includegraphics[width=.8\linewidth]{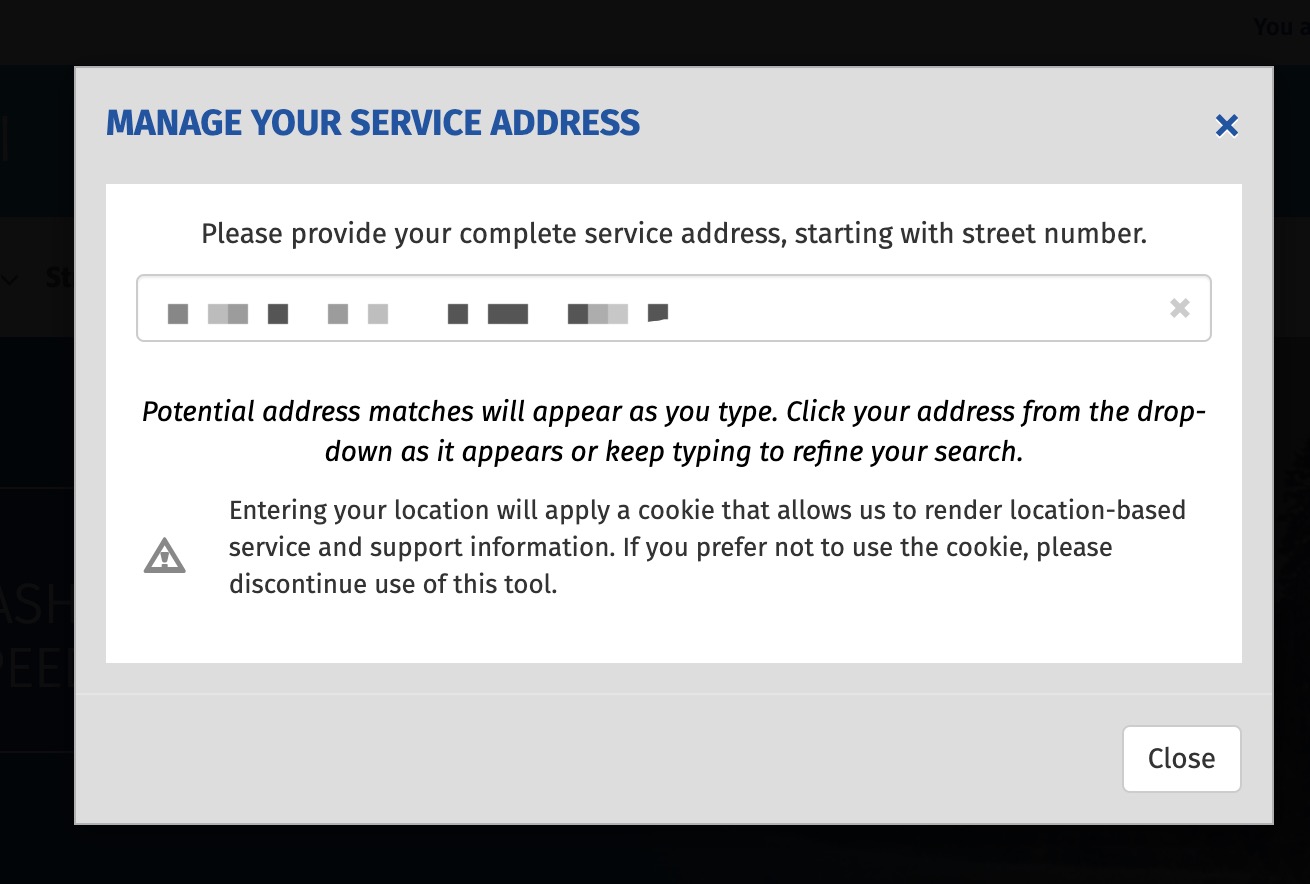}
    \caption{Without an address suggestion.}
    \label{fig:Consolidated Communication No Address Suggestion}
\end{subfigure}
\hspace{0.8in}
\begin{subfigure}[b] {0.4\linewidth}
    \includegraphics[width=.9\linewidth]{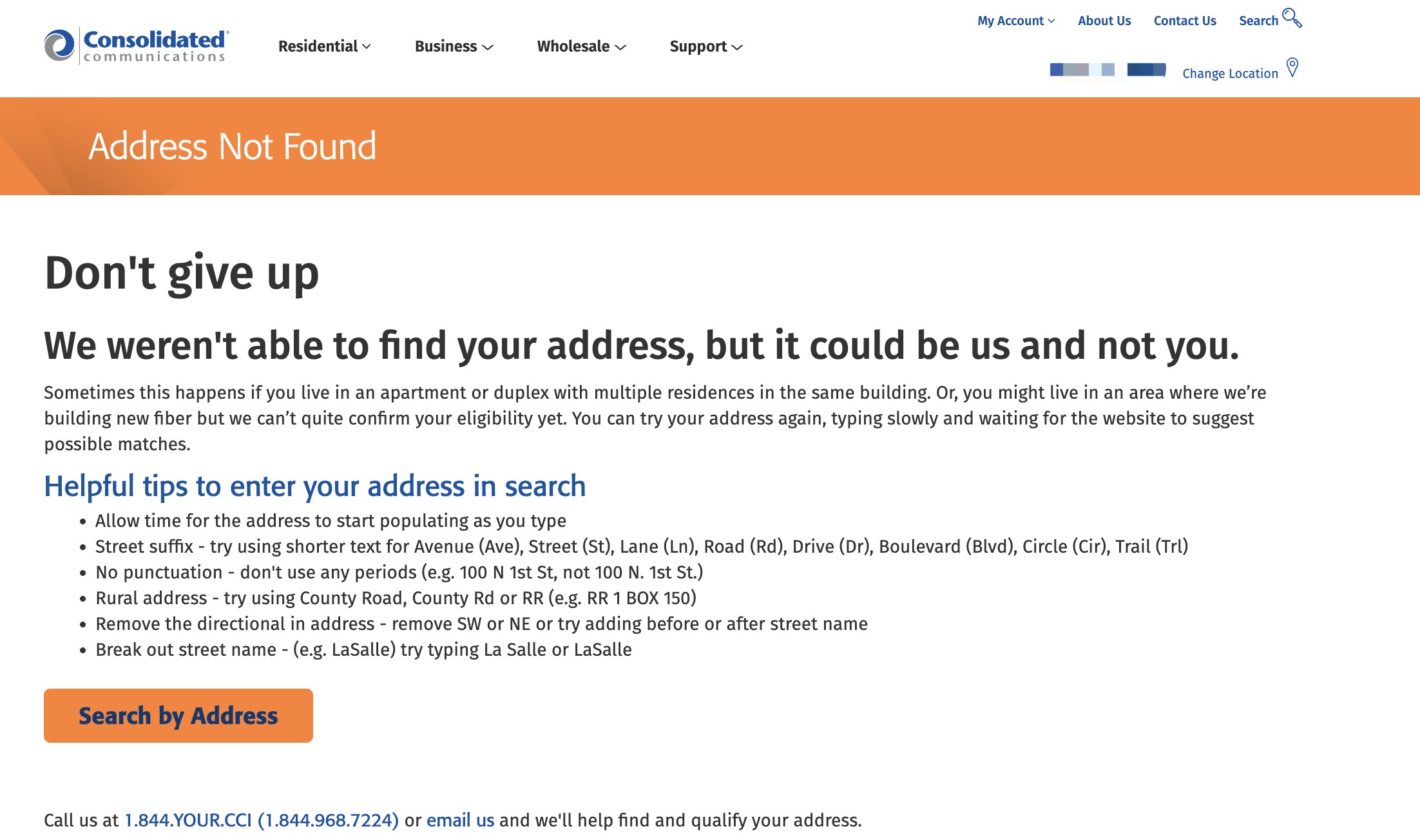}
    \caption{Displaying ``address not found."}
    \label{fig:Consolidated Communication Address Not Found}
\end{subfigure}
\hspace{0.8in}
\begin{subfigure}[b] {0.4\linewidth}
    \includegraphics[width=.9\linewidth]{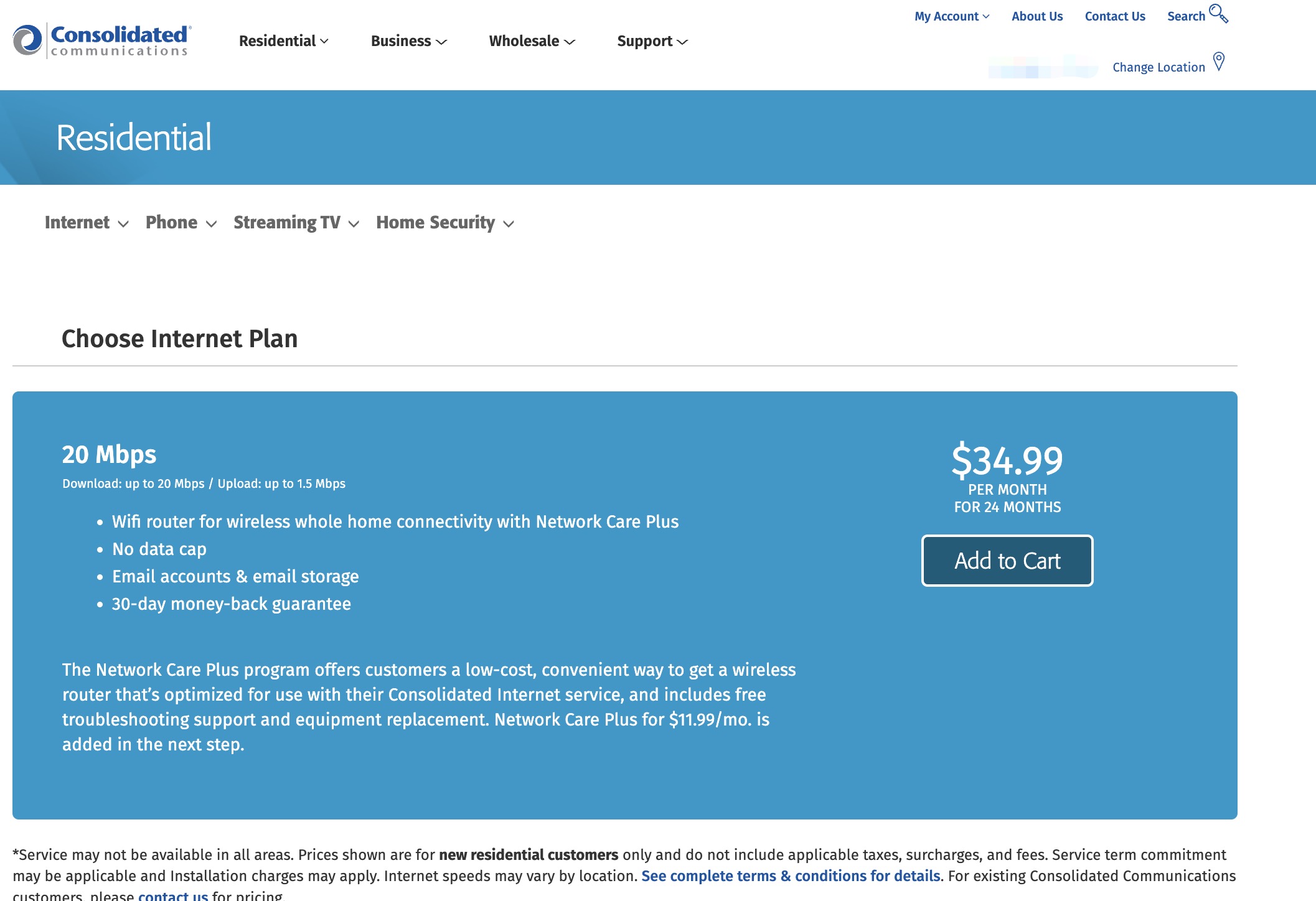}
    \caption{Displaying plan for an existing user.}
    \label{fig:Consolidated Communication Existed Plan}
\end{subfigure}
\hspace{0.8in}
\begin{subfigure}[b] {0.4\linewidth}
    \includegraphics[width=.9\linewidth]{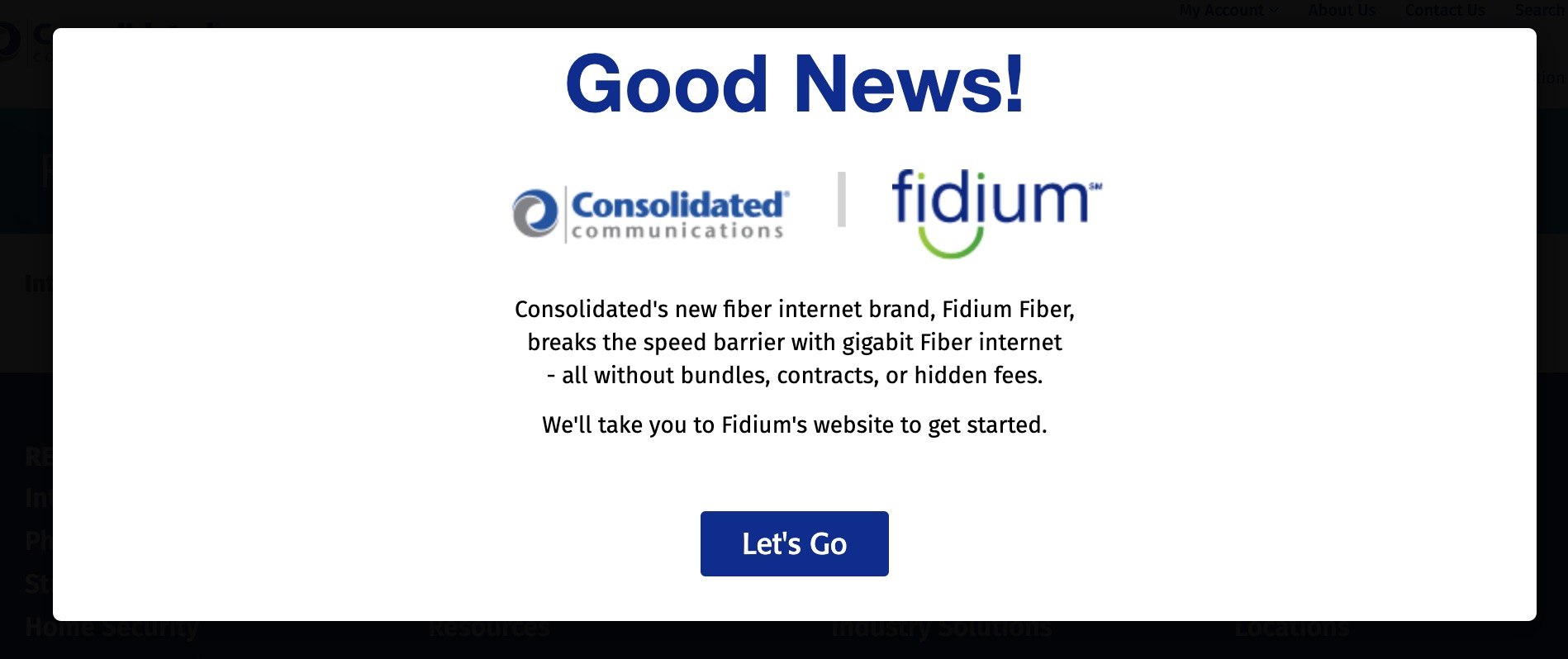}
    \caption{Redirection to Fidium Fiber webpage.}
    \label{fig:Consolidated Communication Redirects to Fidium Fiber webpage}
\end{subfigure}
\hspace{0.8in}
\begin{subfigure}[b] {0.4\linewidth}
    \includegraphics[width=.9\linewidth]{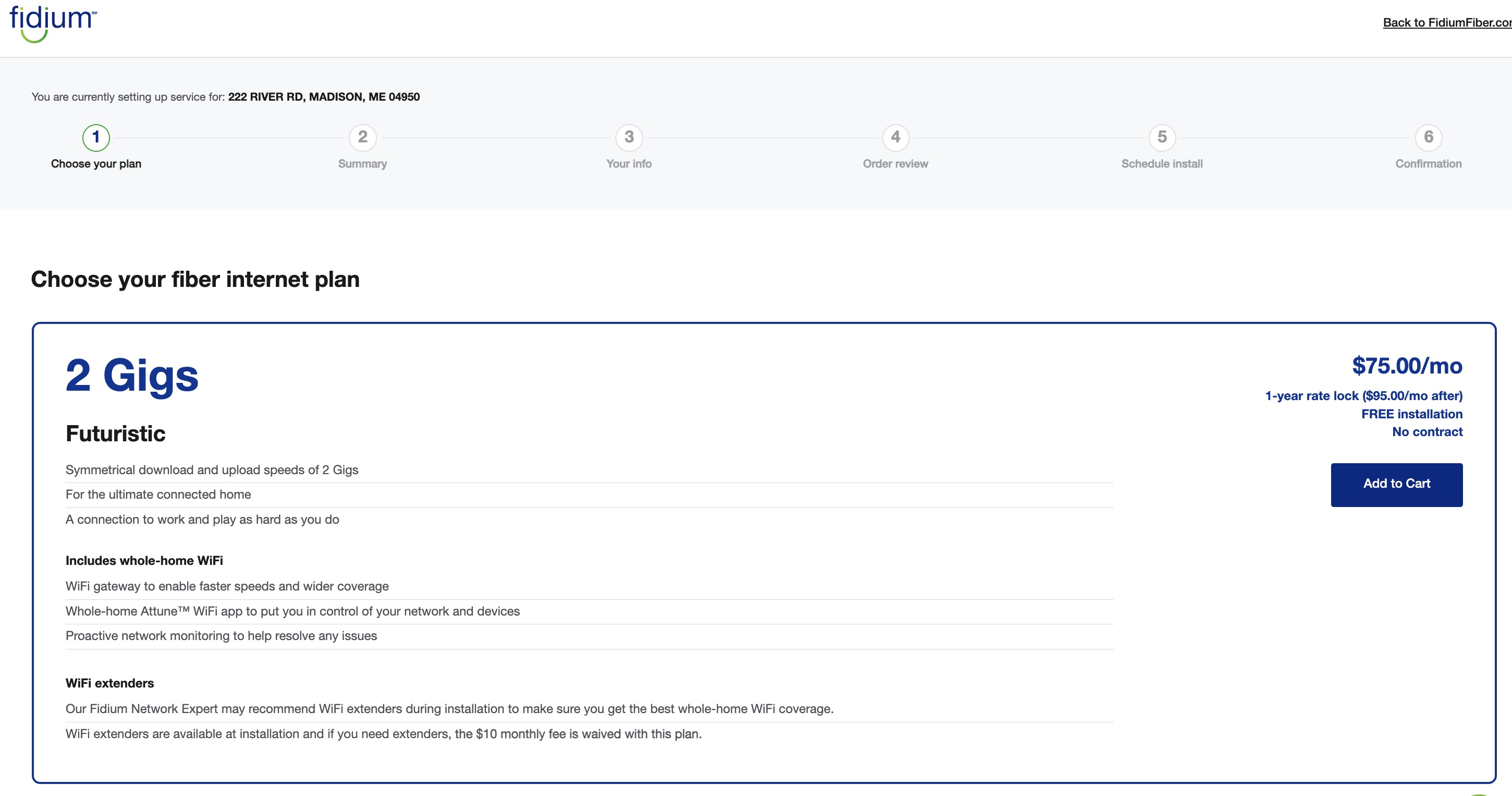}
    \caption{Displaying available plans.}
    \label{fig:Consolidated Communication Fidium Fiber web-page displaying available plans}
\end{subfigure}

\caption{Example webpages from the Consolidated Communication querying process.}
\label{fig:consolidated-query-process}

\end{figure*}

\smartparagraph{Consolidated Communications:}
Consolidated Communications offered service to CAF addresses in 5 of the 15 states in our study. To determine service availability and compliance, we again used BQT to query Consolidated’s website. As with other ISPs, we entered the complete address, and formatted our entry according to the address suggestions provided by the website's ``address look-up tool", as shown in  Figure~\ref{fig:Consolidated Communication Address Suggestion Format}. Finally, we selected the address while it was dynamically resolved by the website's dropdown webform element.
In some cases, Consolidated already had an active subscriber at the queried address. In this case, it asked whether the user would like to view available offers, modify service, or search for a new plan (pictured in Figure~\ref{fig:Consolidated Communication Already Provide Service Modal}).  We selected the latter 
in order to retrieve available plan options for that address. We then logged the address as ``Serviceable" and collected plan information.

In cases where there was no active subscriber, Consolidated's website returned a webpage indicating that service plans are available at that address, as shown in Figure~\ref{fig:Consolidated Communication New Plan}. In such cases, we again logged the address as ``Serviceable" and collected plan information. In other instances, Consolidated Communications offered service through its ``Fidium'' branded service and redirected the user to a Fidium-specific website. In such cases, we also logged the address as ``Serviceable'' and collected information about the Fidium-branded plans available. We observed two other responses from Consolidated Communications's website, neither of which (unlike other ISPs) explicitly indicated that an address was not serviceable. In the first, the dropdown webform element of Consolidated's ``address look-up tool'' failed to provide any address suggestions, accompanied by a modal lacking interactive components for subsequent steps (shown in Figure~\ref{fig:Consolidated Communication No Address Suggestion}). We logged the serviceability status at these addresses as ``Unknown''. Alternately, Consolidated's website would indeed resolve the address correctly but then redirected BQT to a webpage indicating that the address could not be found, as shown in Figure~\ref{fig:Consolidated Communication Address Not Found}. In this case, we logged the serviceability status as ``Address Not Found," and treated the address as if it was not serviceable in our analysis.

Significant updates to the Consolidated Communications website in late November resulted in three distinct web pages for different addresses, this necessitated the meticulous handling of each case. For the majority of addresses, comprehensive details of the coverage plan were prominently displayed on the main website. 
Additionally, for addresses with existing plans, a secondary version of the web page was presented after clicking the ``view offers" button (Figures~\ref{fig:Consolidated Communication Already Provide Service Modal} and~\ref{fig:Consolidated Communication Existed Plan}). For the remaining addresses, after input into the ``address lookup tool", we were redirected to the Fidium Fiber (Consolidated Communications' new fiber broadband service) plan purchasing page (Figure~\ref{fig:Consolidated Communication Redirects to Fidium Fiber webpage}). There, the user could view the broadband plans above 1 Gig (Figure~\ref{fig:Consolidated Communication Fidium Fiber web-page displaying available plans}). If the address was serviced by Fidium, these details were logged as normal. This approach was consistent with our treatment of similar responses from other ISPs.


\smartparagraph{ISPs not funded by CAF: Xfinity and Spectrum.}
Xfinity and Spectrum are two ISPs that do not receive funding from the CAF program.  However, the BQT tool supports querying plans and pricing information from these ISPs, so we include them in the analysis of our third question. 
The BQT tool interacts with the dropdown menus on the Xfinity and Spectrum websites by entering input street addresses and attempting to select the required address from the address suggestions in the dropdown menu. 
In most cases, when an address suggestion is chosen from the dropdown menu, Xfinity and Spectrum explicitly display whether that address is already serviced by them, whether their broadband service is available at that address, or whether that address is out of their service footprint. In the first two cases, we recorded the address as ``Serviceable" and collected plan information. In the third case, we recorded the address as having ``No Service," and excluded it from our analysis. 
In some instances, when an address is entered in BQT, the dropdown menu does not display any address suggestion, and, subsequently, the plan information cannot be obtained for that address. We considered such queries unsuccessful, recorded the result as ``Unknown'', and again excluded the address from our analysis. Finally, for some addresses the ISP website successfully resolved the address but then returned a webpage indicating that the address is invalid. In this case, we logged the address as not serviced, consistent with our approach for other ISPs.

\end{sloppypar}
\end{document}